\documentclass[11pt]{article}
\usepackage{graphicx}
\usepackage{amstext}
\usepackage{caption}
\usepackage{etoolbox}
\usepackage{makeidx}
\include{epsf}
\usepackage{amsfonts}
\usepackage{authblk} 
\usepackage{sectsty}
\usepackage{amsmath,amssymb,epsfig}
\usepackage{amscd}
\usepackage{amsthm}
\usepackage{mathrsfs}
\usepackage{dsfont}
\usepackage[applemac]{inputenc}
\usepackage[english]{babel}
\usepackage{enumitem} 
\usepackage[]{latexsym}
\usepackage{braket}
\usepackage{caption}
\usepackage{hyperref}
\hypersetup{
pdftitle={},%
pdfauthor={},%
pdfsubject={},%
pdfkeywords={},%
colorlinks=true,%
linkcolor=blue,%
citecolor=red,%
linktocpage=true,%
pageanchor=true
}

\newcommand{\be}{\begin{equation}}
\newcommand{\ee}{\end{equation}}
\def\beqa{\begin{eqnarray}}
\def\eeqa{\end{eqnarray}}
\def\bean{\begin{eqnarray*}}
\def\eean{\end{eqnarray*}}
\def\nn{ }

\newcommand{\R}{\mathbb{R}}
\newcommand{\C}{\mathbb{C}}

\newcommand{\tg}{{\tilde{g}}}
\newcommand{\te}{{\tilde{e}}}

\newcommand{\tI}{{\tilde{I}}}
\newcommand{\tJ}{{\tilde{J}}}
\newcommand{\tX}{{\tilde{X}}}
\newcommand{\tY}{{\tilde{Y}}}
\newcommand{\tA}{{\tilde{A}}}
\newcommand{\tK}{{\tilde{K}}}
\newcommand{\bI}{{\mathbf{I}}}
\newcommand{\bJ}{{\mathbf{J}}}
\newcommand{\bX}{{\mathbf{X}}}
\newcommand{\bQ}{{\mathbf{Q}}}

\newcommand{\talpha}{{\tilde\alpha}}
\newcommand{\dd}{{\mathrm{d}}}
\newcommand{\h}{{\underset{\scriptscriptstyle H}{*}}} %Hodge star

\newcommand{\eqn}[1]{(\ref{#1})}
\newcommand{\del}{\partial}
\newcommand{\Tr}[1]{\:{\rm Tr}\,#1}

\renewenvironment{thebibliography}[1]
         {\section*{References}\frenchspacing\small
          \begin{list}{[\arabic{enumi}]}
         {\usecounter{enumi}\parsep=2pt\topsep 0pt
         \settowidth{\labelwidth}{[#1]}
         \leftmargin=\labelwidth\advance\leftmargin\labelsep
         \rightmargin=0pt\itemsep=1pt\sloppy}}{\end{list}}

 \numberwithin{equation}{section}

\title{\textbf{ T-Dualities and Doubled Geometry of  the Principal Chiral Model}\vspace{0.5cm}}
\date{}

\author[1]{Vincenzo E. Marotta}
\author[2]{Franco Pezzella}
\author[2,3]{Patrizia Vitale}
\affil[ ]{}
\affil[1]{\textit{\footnotesize Department of Mathematics, 
Heriot-Watt University
Colin Maclaurin Building, Riccarton, Edinburgh EH14 4AS, U.K.}}
\affil[2]{\textit{\footnotesize INFN-Sezione di Napoli, Complesso Universitario di Monte S. Angelo Edificio 6, via Cintia, 80126 Napoli, Italy.}}
\affil[3]{\textit{\footnotesize Dipartimento di Fisica ``E. Pancini'', Universit\`a di Napoli Federico II, Complesso Universitario di Monte S. Angelo Edificio 6, via Cintia, 80126 Napoli, Italy.}}
\affil[ ]{}
\affil[ ]{\footnotesize e-mail: \texttt{vm34@hw.ac.uk, franco.pezzella@na.infn.it, patrizia.vitale@na.infn.it}}
\begin{document}
\maketitle
\begin{abstract}
\small The Principal Chiral Model  (PCM) defined on the group manifold of $SU(2)$ is here investigated with the aim of getting a further deepening of its relation with Generalized Geometry and Doubled Geometry.  A one-parameter family of equivalent Hamiltonian descriptions is analysed, and cast into the form of Born geometries. Then  $O(3,3)$ duality transformations  of the target phase space are performed and we show that the resulting dual models are defined on the group $SB(2,\C)$ which is  the Poisson-Lie dual  of $SU(2)$ in the Iwasawa decomposition of the Drinfel'd double $SL(2, \mathbb{C} )$.    
 A parent action with doubled degrees of freedom and configuration space $SL(2, \mathbb{C})$ is then defined that reduces to either one of the dually related models, once suitable constraints are implemented. 
 
\end{abstract}

\newpage
\tableofcontents

\section{Introduction}

Duality symmetries play a fundamental role in String Theory since they provide a powerful tool for investigating the structure of the target spacetime from the string point of view by relating, in the usual sigma-model approach, backgrounds which otherwise would be considered different. The Abelian T-duality \cite{ giveon94, alvarez95, duff} (where T stands for Target-space) is  a well-known example of them.  It is a distinctive symmetry of strings since, differently from particles, one-dimensional objects can wrap $d$ non-contractible cycles. This implies the presence of winding modes $w^{a}$ $(a=1, \dots, d)$ that have to be added to the ordinary momentum modes $p_{a}$ which take integer values along compact dimensions. On a $d$-torus $T^{d}$, the Abelian T-duality is  an $O(d,d; {\mathbb Z})$ string symmetry under, roughly speaking, the mapping of the radii of the compact dimensions into their inverse, together with the exchange of momentum and winding modes: in this way it establishes a connection between two apparently different but dual target spacetimes. From the sigma model point of view, the necessary condition to work out a dual to some background was, initially,  that the latter possess an Abelian group of isometries \cite{buscher1,buscher2, rocekverlinde} excluding in this way many physically relevant classical string vacua from being considered.

After the work in ref. \cite{ossa-quevedo}, it was understood that T-duality symmetries could also be associated with the non-Abelian isometries of the target manifold and, subsequently, the notion of Abelian and non-Abelian T-duality was extended to the one of Poisson-Lie T-duality \cite{klimcik1, klimcik2, klimcik3}. Briefly, the term Abelian T-duality refers to the presence of global Abelian isometries in the target spaces of both the paired sigma models; non-Abelian T-duality refers to the existence of a global Abelian isometry on the target space of one of the two sigma-models and of a global  non-Abelian isometry on the other. Finally,  the Poisson Lie T-duality generalizes the previous definitions to all the other cases, including the one of a pair of sigma models both having non-Abelian isometries in their target spaces.

Beyond the string world-sheet action, a category  of models that reveal themselves to be very helpful in understanding the above mentioned T-dualities is provided by sigma models whose target configuration space is a Lie group $G$ with  $\mathfrak{g}$ its Lie algebra. These are the so-called  Principal Chiral Models (PCM). Studying these models has led to abandoning the requirement of the existence of isometries for the target space as the condition for the existence of dual counterparts.  Indeed, the relevant structure in this case reveals to be the one of Drinfel'd double for $G$ together with the well-established notion of Poisson-Lie symmetries \cite{drinfeld1}-\cite{ kossmann}.  The Drinfel'd double of a Lie group $G$ is defined as a Lie group $D$, with  dimension twice the one of $G$, such that its Lie algebra $\mathfrak{d}$ can be decomposed into a pair of maximally isotropic sub-algebras, $\mathfrak{g}, \tilde {\mathfrak{g}}$  with respect to a non-degenerate invariant bilinear form on $\mathfrak{d}$, with $\mathfrak{g},   \tilde{\mathfrak{g}}$, respectively the Lie algebra of $G$ and its dual algebra.\footnote{  An isotropic subspace of the Lie algebra $\mathfrak{d}$ is such that the bilinear form evaluated on any couple of   vectors lying in that subspace vanishes; maximally isotropic means that the subspace cannot be enlarged while preserving the property of isotropy.} The dual algebra is endowed with a Lie bracket which has to be compatible with  existing  structures,  in a precise sense which will be clarified below.   Any such triple,  $(\mathfrak{ d}, \mathfrak{ g},  \tilde{\mathfrak{g}}) $, is referred to as a Manin triple. By exponentiation of $ \tilde{\mathfrak{g}}$ one gets the dual Lie group $\tilde{G}$ such that locally  $ D\simeq G\times \tilde{ G}$.  
The simplest example is the cotangent bundle of any $d$-dimensional Lie group $G$, $T^*G\simeq G \ltimes \R^d$, which  we shall call the  classical double,    with trivial Lie bracket for the dual algebra $ \tilde{\mathfrak{g}}\simeq \R^d$. 
 For every  decomposition  of the Drinfel'd double  $D$ into dually related  subgroups $G, \tilde{ G}$,  it is possible to define a couple of PCM's having as  target configuration space either of the two subgroups. Hence, every PCM has its dual counterpart for which the role of $G$ and its dual $\tilde{G}$ is interchanged. The set of all decompositions of $\mathfrak{d}$ into maximally isotropic subspaces (not necessarily subalgebras),   plays the role of the modular space of sigma models mutually connected by an $O(d,d)$ transformation. In particular, for the manifest Abelian T-duality of the string model on the $d$-torus, the Drinfel'd double is $D=U(1)^{2d}$ and  its modular space,  is  in one-to-one correspondence with $O(d,d; \mathbb{Z})$ \cite{klimcik3}.

In this paper, we are going to show that the target phase space of the $SU(2)$ PCM can actually be replaced by the Drinfel'd double of $SU(2)$, namely the group $SL(2,\C)$,  {\it without modifying the dynamics}. This observation, based on previous work by  Rajeev \cite{R89, R892}, is the main motivation for our interest in the model, since it allows to discuss Poisson-Lie T-duality, a   generalization of    Abelian T-duality, in  a situation where it is a true symmetry of the undeformed dynamics. Later on  we shall discuss more in detail this important point. 

Non-linear sigma models  have been investigated in relation to Poisson-Lie duality, with or without reference to string theory, by many authors (see for example  \cite{Sfetsos:1998}-
%, Falceto:2001,Calvo:2003, Sfetsos:2009, Severa:2017,Hassler:2017,Jurco:2017,
\cite{Chatzistavrakidis:2018} and ref.s therein). A closer approach to the one which will be pursued in the paper has been adopted in    \cite{sfetsos}, \cite{edwards:nonpub}.

 Another motivation for analyzing sigma models having as target phase-space a Drinfel'd double Lie group $D$,   consists in the fact that it allows to   establish some connections with  Generalized Geometry (GG), by virtue of the fact that tangent and cotangent vector fields of the group manifold $G$ may be  respectively related to  the span of the Lie algebra $\mathfrak{g}$ and its dual, $\tilde{\mathfrak{g}}$. (Let us briefly recall that  GG     \cite{hitchin1, hitchin2, gualtieri:tesi} contains, roughly speaking,  two main ingredients - the first consists in replacing the tangent bundle $T$ of a manifold $M$ with  $T\oplus T^*$, a bundle with the same base space $M$ but  fibers given by the direct sum of tangent and cotangent spaces, and the second in replacing the Lie bracket among sections of $T$, that is vector fields, with the Courant bracket which involves vector fields and one-forms.) Moreover, Doubled Geometry (DG) may play a role in describing the generalized dynamics on the tangent bundle $TD \simeq D \times  \mathfrak{d}$, which we shall do in order to describe within a single action both dually related models. 
  Both  GG and DG have revealed to be very suitable in describing the geometry of Double Field Theory (DFT) \cite{HZ}-\cite{nunez}. 
DFT provides a proposal to incorporate the Abelian T-duality of a compactified string on a $d$-torus $T^{d}$ in a $(\mathcal{G},B)$-background as a manifest symmetry of the string effective field theory.  More precisely, DFT  is supposed to be an $O(d,d; \mathbb{Z})$ manifest spacetime effective field theory that should derive from a manifestly T-dual invariant formulation of a string world-sheet action in which T-duality is made manifest. Such a formulation was proposed in ref.s \cite{duff, tseytlin1, tseytlin2} and later developed in ref.s \cite{hull}-\cite{copland2} (see more recent works in \cite{pezzella1}-\cite{ma-pezzella}). This string action has to contain information about windings and therefore it is based on two sets of coordinates: the usual string coordinates $x^{a}(\sigma, \tau)$ in the target space, having the momenta $p_{a}$ as conjugate, and the dual coordinates, $\tilde{x}_{a} (\sigma, \tau)$ having the winding modes as conjugate momenta. In this way the $O(d,d; \mathbb{Z})$ duality results to be a manifest symmetry of the world-sheet action even paying the price of loosing the manifest covariance in the world-sheet two dimensions.  A doubling of all the $N$ spacetime degrees of freedom in the low-energy effective action first occurred in ref.s \cite{siegel1}-\cite{siegel4} where a manifestly $O(N,N; \mathbb{R})$ effective action in the target space was obtained, and such symmetry was realized linearly, loosing this time the manifest Lorentz invariance in the target space.   In order to understand the role of Doubled and Generalized Geometries in a simpler context, the doubling of degrees of freedom has been analyzed in the context of  finite-dimensional dynamical systems, such as the dynamics of  a charged particle in presence of a uniform distribution of magnetic monopoles,  in  ref.s \cite{Bakaslust, kupr, Szabo:2019}, where the doubling is justified by the otherwise violated Jacobi identity for the algebra of observables \footnote{Violation of Jacobi identity can be related to the violation of associativity of the star product of the quantized theory \cite{jackiw}. See refs. \cite{CGLMV}-\cite{GLVV} in relation to the problem of finding an associative star product for the electron-monopole system and related problems.}. Moreover,  it is worth mentioning  that the occurrence of auxiliary degrees of freedom  is also typical of   other geometric theories, such as those based on   Noncommutative Geometry. Noncommutative gauge theories require  that the gauge group be enlarged (see for example \cite{szabonc} for a review). The differential calculus itself may be bigger than in the commutative case (see  \cite{MVZ, MVZ2} for an example  in three dimensions and \cite{MVW} for an application to two-dimensional gauge theory). 
Renormalizability of noncommutative field theories entails  the introduction of auxiliary parameters, such as  for the Grosse-Wulkenhaar model \cite{GW}, or the translation-invariant model \cite{tinv,tinv2}. Last but not least, noncommutative extensions of Palatini-Holst theory of gravity  imply the doubling of the tetrad degrees of freedom, leading to a bi-tetrad theory of gravity, with the manifestation of  new  duality symmetries \cite{dCSV18, aschiericastellani}.

It should be clear that models whose carrier space of the dynamics is the manifold of a  Lie group  can be very helpful in better
understanding T-duality and doubling of the degrees of freedom. The latter are naturally described in the framework outlined above,  by generalizing the dynamics   originally defined on $G$ to a dynamics on the Drinfel'd double $D$, through the introduction of a natural parent  action;  T-duality is naturally provided by the exchange of the two partner groups $G$ and  $\tilde{G}$.   The formulation of Double Field Theory on group manifolds,  including its relation with Poisson-Lie symmetries,   has been studied in ref.s \cite{blumen1, blumen2}. For recent results see ref.s \cite{Demulder:2018, Mori:2019}.

This is the second of a series of two papers. In the first one \cite{MPV18}, we have studied the three-dimensional isotropic rigid rotator (IRR) that provides the simplest one-dimensional sigma model having ${\mathbb R}$ as a source space and the group manifold $SU(2)$ as target configuration space. We have then introduced a model with target space  the dual group $SB(2,{\mathbb C})$ and considered the symmetry properties of the two models  within an extended model on the Drinfel'd double $SL(2,  {\mathbb C})$, formulated in terms of a {\em parent action}. In particular, we have emphasized how a natural para-Hermitian structure emerges on the Drinfel'd double and can be used to provide a doubled formalism for the pair of theories.  
The IRR model is too simple to exhibit symmetry under duality transformation, being a $0+1$ field theory but it has paved the way for a genuine 1+1 field theory, the SU(2) Principal Chiral Model which, while being modeled on the IRR system, certainly exhibits interesting properties under duality transformations: therefore, the $SU(2)$ Principal Chiral Model is the topic of this second paper. 

More precisely, we elaborate on an old intuition due to S. G. Rajeev which dates back to the 80's \cite{R89, R892} where  the principal $SU(2)$ chiral model is shown to exhibit a whole one-parameter family of alternative Hamiltonians and alternative Poisson algebras, all equivalent from the point of view of the dynamics [also see  ref. \cite{RSV93} where the construction is extended to the Wess-Zumino Witten model,  and  ref. \cite{RSV96}  where the integrability is analyzed in terms of Lax pairs].  The model is described in the Hamiltonian approach by a pair of fields $J^i(t,\sigma), I_i (t,\sigma)$, the so-called currents, which are valued in the phase space $T^*SU(2)$, that we shall refer to as the {\it target phase space}. Let us briefly recall that, topologically, $T^*SU(2)$ is the manifold $S^3\times \R^3$, while, as a group, it is the semidirect product of $SU(2)$ with the Abelian group $\R^3$. As a Poisson manifold it is known to be symplectomorphic to the group $SL(2,C)$\footnote{when endowed with  appropriate Poisson brackets \cite{marmoibort}}, which should come with no surprise since the two have the same topology. Last but not least, $T^*SU(2)$ and $SL(2,\C)$  are both Drinfel'd doubles of the group $SU(2)$ \cite{drinfeld1}-\cite{kossmann}. The former, which we shall call {\it classical double},   is the trivial one, with  Abelian algebra of momenta and can be obtained from the latter via group contraction. 

The many different geometric structures which are compatible with the manifold $S^3\times \R^3$ will play a   crucial role all over the paper.  To start with, the whole construction relies on the generalization of the affine algebra of currents,  associated with  the semi-direct sum $\mathfrak{su}(2)(\R)\ltimes \mathfrak{a}(\R)$,   being $ \mathfrak{a}(\R)$ an Abelian Lie algebra, to a fully semi-simple Kac-Moody algebra which is  either $\mathfrak{su}(2)(\R)\oplus \mathfrak{su}(2)(\R)$ or $\mathfrak{sl}(2,\C)(\R)$. Here by $\mathfrak{g}(\R)$ we shall indicate the affine algebra associated to the Lie algebra $\mathfrak{g}$. Interestingly, this construction can be understood in terms  of Born Geometry \cite{Borngeomf}-\cite{marottaszabo}, which we shortly review and adapt to our model.
By slightly generalizing the Poisson Kac-Moody algebra with the introduction of a second parameter, and performing an $O(3,3)$ transformation over the target phase space, we show that a family of sigma models with target configuration space the group manifold of $SB(2,\C)$ is obtained, which deserves the name of T-dual models.  Moreover,  the vanishing value of one  of the two parameters  corresponds to the original $SU(2)$ PCM with canonical splitting of its current algebra, whereas the vanishing of the remaining parameter correctly reproduces the dual current algebra $\mathfrak{sb}(2,C)(\R) \ltimes \mathfrak{a}(\R)$, but the Hamiltonian exhibits a singular behaviour which is yet to be understood. 
%the Lagrangian submanifolds corresponding to $SU(2)$ and $SB(2, \mathbb{C})$ are found  for two opposite values of that parameter, 
%leading to an interesting generalization of T-duality. 

Let us stress  here that the one-parameter family of Hamiltonian models, re-proposed in Eqs. \eqn{modipo1}-\eqn{modiha}, but already contained in \cite{R89, RSV93}, yields an {\it equivalent} description of the standard dynamics of the PCM. Namely, for each value of $\tau$ the dynamics is one and the same, up to rescaling  the fields by appropriate factors of $\tau$. In this sense, it is different from the deformations introduced in ref. \cite{klimcikdefo}, which are {\it true} deformations of the dynamics. We prove explicitly that the same result holds for the two-parameter generalization represented by the algebra \eqn{IIcural}-\eqn{IKcural}, upon rescaling and   linear transforming $(I,K)\rightarrow (I,J)$. 
%Non-linear sigma models  have been investigated in relation to Poisson-Lie duality by many authors \cite{severa, vysoky, falceto, ..}. A close approach to ours can be found %in in   \cite{sfetsos}, \cite{edwards:nonpub}.
%The relation of the above mentioned generalization with Manin triples and Drinfel'd doubles was already suggested in ref. \cite{RSV96} and has been fully explored in  ref. \cite{MPV18}  for the IRR, i.e. the $0+1$-dimensional analogue of the principal chiral model.

The paper is organized as follows. 

In Section \ref{IRR} the results obtained in ref. \cite{MPV18} are reviewed  for the  isotropic rigid rotator thought of  as a dynamical model over the group manifold $SU(2)$ with a  dual partner defined on the  dual group  $SB(2, \mathbb{C})$. The two groups appear in the Iwasawa decomposition of the Drinfel'd double $SL(2, \mathbb{C})$ whose structure is recalled together with the one of its  Lie algebra  $\mathfrak{sl}(2,\C)\simeq \mathfrak{su}(2) \Join \mathfrak{sb}(2,\C)$. 

In Section \ref{PCM} the generalization of the dynamics of the rigid rotor to the $SU(2)$ Principal Chiral Model is described in the Lagrangian and Hamiltonian approach, with the introduction of the Poisson  algebra of currents, which is the affine algebra $\mathfrak{su}(2)(\R)\ltimes \mathfrak{a} (\R)$. The existence of a whole one-parameter family of alternative Hamiltonians with a fully semi-simple affine algebra $\mathfrak{sl}(2,\C)(\R)$ is discussed and its interpretation in terms of Born geometries is analyzed. 

In Section \ref{PLD} a family of T-dual models is introduced in the Hamiltonian formalism and it is shown that the target configuration space for the latter is the group manifold $SB(2,\C)$. In Subsection \ref{duallagsec} a different perspective is adopted. Analogously to what has been done for the Isotropic Rigid Rotator, a natural Lagrangian model is constructed directly on the dual group $SB(2,\C)$ and its relation to the dual models introduced previously is analyzed. %A new kind of duality emerges in this context. 

Finally, in the spirit of Double Field Theory, in order to build a model where the symmetries exhibited by the dynamics are manifest, a parent action is constructed in Section \ref{DPCM} having as target configuration space  the Drinfel'd double $SL(2,\C)$, hence doubling the degrees of freedom.   From it, either of the dual partner models can be recovered, by gauging one of its global symmetries.  

Conclusions and Outlook are reported in the final Section \ref{conclus}. An Appendix follows where the current algebras for all models considered are explicitly derived.

\section{The Isotropic Rigid Rotator}\label{IRR}

In this Section we shortly review the results obtained in  ref. \cite{MPV18}   for the isotropic rigid rotator as a dynamical model over the group manifold of $SU(2)$, and its dual model having as configuration space the group manifold of the Lie-Poisson dual of $SU(2)$, the group  $SB(2,\C)$. 
Moreover, we  briefly recall the Drinfel'd double structure of the group $SL(2,\C)$   and the bialgebra nature of its Lie algebra $\mathfrak{sl}(2,\C)\simeq \mathfrak{su}(2) \Join \mathfrak{sb}(2,\C)$. 

The classical action which describes the dynamics can be chosen to be:
\be
S_0= -\frac{1}{4} \int_\R \Tr \bigl( \phi^*[ g^{-1} \mathrm{d} g]\wedge \h \phi^*[g^{-1} \dd g] \bigr) =-\frac{1}{4}\int_\R  \Tr (g^{-1}\frac{d g}{dt})^2 dt \label{lag}
\ee
with $\phi  :t\in \R\rightarrow g\in  SU(2)$,    $\h$ the Hodge  
star operator on the source space $\R$,  $\h dt = 1$,   $\Tr$ the trace over the Lie algebra and 
$g^{-1} d g$  the Maurer-Cartan left-invariant  one-form on the group manifold.  With an abuse of notation, the pull-back map $\phi^*$ will be omitted since now on. 

Therefore the model can be regarded as a 
  $(0+1)$-dimensional, group valued, field theory. 
  
 In order to motivate the interest for such a model,  it is worth anticipating here that, with 
 $g:\R^{1,1}\rightarrow SU(2)$ and  $\R^{1,1}$ the Minkowski spacetime, the action \eqn{lag} generalizes to the one describing  the Principal Chiral Model, that is to say, a non-linear sigma model with target space the group manifold of $SU(2)$. 
 
 By choosing the parametrization 
  $g= y^0 \sigma_0 
+i  y^i {\sigma_i}$, with $(y^0)^2+ \sum_i (y^i)^2=1$ and $\sigma_0$ the identity matrix, $\sigma_i$  Pauli matrices and the inverse relations   
$$
y^i=-\frac{i}{2} \Tr g\sigma_i, \; \; y^0= \frac{1}{2} \Tr g \sigma_0, \;\;\; i=1,..,3 \,\,\,\, ,
$$
one has:
\be
g^{-1} \dot g=i (y^0\dot y^i-y^i \dot y^0+  {\epsilon_{jk}}^i {y^j \dot y^k })\sigma_i  := i \dot Q^i\sigma_i
\ee 
being
\be
\dot Q^i= y^0\dot y^i-y^i \dot y^0+  {\epsilon_{jk}}^i {y^j \dot y^k }
\ee
the {\it left generalized velocities.}\footnote{Had we chosen to work with the right-invariant Maurer-Cartan one-form we would have introduced {\it right generalized  velocities.}}
The 
 Lagrangian  then  reads as:
\be
\mathcal{L}_0=\frac{1}{2} (y^0\dot y^j-y^j \dot y^0+  {\epsilon_{kl}}^j y^k \dot y^l)(y^0\dot y^r-y^r \dot y^0+  
{\epsilon_{pq}}^r y^p \dot y^q)\delta_{ir} := \frac{1}{2} \dot Q^j \dot Q^r \delta_{j r}.
\ee
This 
yields the following equations of motion:
\be
{\sf L}_{\Gamma} \dot Q^i = 0,\;\;\;\; {\rm or~equivalently~~} {\sf L}_\Gamma \left(g^{-1}\frac{dg}{dt}\right)=0
\ee
with ${\sf L}_\Gamma$ the Lie derivative with respect to $\Gamma=\frac{d}{dt}$. 

The cotangent bundle (left) coordinates are represented by  $(Q^i, I_i)$  with $I_i$ being the left conjugate momenta:
\be
I_i= \frac{\del {\mathcal L}_0}{\del \dot Q^i}= \delta_{i j} \dot Q^j .
\ee
The {Hamiltonian} is thus
$
\mathcal{H}_0= \frac{1}{2}I_i I_j \delta^{ij} 
$
with Poisson brackets (see ref. \cite{MPV18} for details) given by:
\beqa
\{y^i,y^j\}&=&0 \label{yy}\\
\{I_i,I_j\}&=&{\epsilon_{ij\;}}^k I_k \label{II}\\
\{y^i,I_j\}&=&\delta^{i}_j y^0 +{\epsilon_{jk}}^i y^k \;\;\;  {\rm or ~ }\;\;\; \{g, I_j\}= i g \sigma_j g
\label{yI}
\eeqa
which lead to the dynamics described by the following equations:
\be
\dot I_i= 0,\;\;\; g^{-1}\dot g= i I_i \delta^{ij}\sigma_j.
\ee
The fiber coordinates  $I_i$ are   associated with the angular momentum components and the base space coordinates $g\equiv (y^0, y^i)$ to the orientation of the rotator. 
\\
As well-known, $I_i$ are constants of the motion, while  $g$ undergoes a uniform precession. 

Remarks:
\begin{itemize}
%\item  The fibers of the tangent bundle $TSU(2)$ are  $\mathfrak{su}(2)\simeq \R^3$, with $\dot Q^i$ denoting vector fields components
%\item The fibers of the cotangent bundle $T^*SU(2)$ are isomorphic to the dual Lie algebra $\mathfrak{su}(2)^*$. Again $\R^3$, but $I_j$ are now components of one-forms. 
\item As a group $T^*SU(2)$ is the semi-direct product $SU(2) \ltimes \R^3$  with Lie algebra $\mathfrak{su}(2) \ltimes \R^3$ and Lie brackets given by:
%$\;$\\
$$
\left[L_i,L_j\right]  =  {\epsilon_{ij}}^k L_k \label{JJ} ~~~~~~
\left[T_i,T_j\right] =  0 \label{PP}~~~~~~
\left[L_i,T_j\right]= {\epsilon_{ij}}^k T_k  \,\,\, .
$$
Here $L_i, T_i, i=1,2,3$ generate respectively the algebra $\mathfrak{su}(2)$ and the algebra  $\R^3$.
\item The  non-trivial Poisson brackets \eqn{yy}-\eqn{yI} are the Kirillov-Souriau-Konstant (KSK) brackets on  the dual algebra $\mathfrak{\tilde{g}}$ .
\end{itemize}
Starting from these remarks, in ref. \cite{MSS92} the carrier space of the dynamics  has been  generalized to   $SL(2, \C)$, the non-trivial Drinfel'd double of $SU(2)$, which, roughly speaking, can be obtained by deforming  the Abelian subgroup $\R^3$ of the semi-direct product above, and  a similar generalization has been proposed  for the Principal Chiral Model  \cite{R89} and  the Wess-Zumino-Witten Model \cite{RSV93}.

The algebra $\mathfrak{sl}(2,\C)$ is usually described in terms of the  generators
$e_i= \sigma_i/2, b_i= i e_i$, $i=1,2,3$,  with Lie brackets 
\be
 [e_i, e_j]= i {\epsilon_{ij}}^k e_k, ~~~~~[e_i, b_j]= i {\epsilon_{ij}}^k b_k, ~~~~~[b_i, b_j]=- i {\epsilon_{ij}}^k e_k \,\,\, .
\ee
It is equipped with two non-degenerate invariant scalar products: 
\be
\Braket{u,v}=2{\rm Im}(\mbox{Tr}(uv)) \quad \forall u,v \in \mathfrak{sl}(2,\mathbb{C})  \label{prod1}
\ee
and 
\be
(u,v)=2{\rm Re}(\mbox{Tr}(uv)) \quad \forall u,v \in \mathfrak{sl}(2,\mathbb{C}). \label{prod2}
\ee
With respect to the first one (the Cartan-Killing metric),  one has two maximal isotropic subspaces, spanned by $\{e_i\}$, and the linear combination
\be \label{etilde}
\tilde e^i= b_i -\epsilon_{ij3}e_j .
\ee
 Indeed the following relations hold:
\be
\Braket{e_i, e_j}= \Braket{\tilde e^i, \tilde e^j}= 0~~~~~~~  \mbox{and}~~~~~~~\Braket{e_i, \tilde e^j}=\delta_i^j .
\ee
The generators $\{e_i\}$,  $\{\tilde e^i\}$ span two non-commuting   subalgebras of $\mathfrak{sl}(2,\C)$ with Lie brackets:
\be
[e_i, e_j]= i {\epsilon_{ij}}^k e_k, ~~~~~[\tilde e^i,  e_j]= i{\epsilon_{jk}}^i \tilde e^k+ i e_k {f^{ki}}_j , ~~~~~[\tilde e^i, \tilde e^j]= i {f^{ij}}_k \tilde e^k. \label{bi-algebra}
\ee
 In particular, $\{\tilde e^i\}$ span the Lie algebra of $SB(2,\C)$, the {\it dual group} of $SU(2)$  with 
\be
{f^{ij}}_k=\epsilon^{ij l}\epsilon_{l3k}. \label{constf}
\ee
Each algebra acts on the other one non-trivially by coadjoint action, as it can be read from Eq. \eqn{bi-algebra} and therefore we denote the total algebra by $\mathfrak{sl}(2,\C) = \mathfrak{su}(2) \Join\mathfrak{sb}(2,\C)$, with $\Join$ generalizing the semi-direct sum. 

Summarizing:
\begin{itemize}
\item 
 $\mathfrak{sl}(2,\C) $  can be endowed with  a Lie bialgebra structure; 
\item
the role of $\mathfrak{su}(2)$ and its dual algebra can be interchanged.
\end{itemize}
 The triple  $(\mathfrak{sl}(2,\mathbb{C}), \mathfrak{su}(2), \mathfrak{sb}(2,\mathbb{C}))$ is called a  {\it Manin triple}.
 
 The construction can be generalized to any Lie group $G$. Given $\mathfrak{d}= \mathfrak{g}\Join \mathfrak{\tilde{g}}$ , the group 
 $D$ with Lie algebra $\mathfrak{d}$ is the Drinfel'd double and $G$, $\tilde{G}$ are dual groups.  
 For ${f^{ij}}_k=0 \;\;\; D \rightarrow T^*G$, while 
for ${c_{ij}}^k=0 \;\; D \rightarrow T^* \tilde{G} $, with  ${c_{ij}}^k$ the structure constants of $\mathfrak{g}$ and ${f^{ij}}_k$  the structure constants of $\mathfrak{\tilde{g}}$. 
Therefore $D$ generalizes both the cotangent bundle of $G$ and of $\tilde{G}$. 

 The bialgebra structure induces Poisson structures on the  group manifold of the double $D$
  which generalize both those of $T^*G$ and of $T^* \tilde{G}$ and reproduce  the KSK brackets on coadjoint orbits of $G$, $\tilde{G}$ when ${f^{ij}}_k=0, {c_{ij}}^k=0$ respectively. 
  %Indeed the Lie algebra structure 
 %$[\;,\;]_{\mathfrak{su}(2)}$ induces a Poisson structure, $\Lambda$, on  the algebra of functions $\mathcal{F}(SB(2,\C))$ and dually the Lie algebra structure $[\;,\;]_{\mathfrak{sb}(2,\C)}$ induces a Poisson structure, $\tilde\Lambda$ on  $\mathcal{F}(SU(2) )$.  
%Let us review   these structures in some detail. 
%The  double group $SL(2,\C)$ can be endowed with PB's which generalize both those of $T^*SU(2)$ and of $T^*SB(2\C)$
For $\gamma\in D$ and being $r = \lambda  \tilde{ e}^i\otimes e_i$ with $\lambda \in \R $ the classical Yang-Baxter matrix, the  brackets 
\be
\{\gamma_1,\gamma_2\}= -\gamma_1\gamma_2 r^* -r \gamma_1\gamma_2 
\ee
where $\gamma_1= \gamma\otimes 1, \gamma_2= 1\otimes \gamma_2$, $r^*= -\lambda  e_i\otimes \tilde {e}^i$, 
%$I_i$ and $J^i$ from the ones stated in \cite{marmo:articolo1}.\\
% $$r =  \tilde{ e}^i\otimes e_i ,~~~~~~r^*= - e_i\otimes \tilde {e}^i $$ is the classical Yang Baxter matrix 
can be shown to define a Poisson structure on the group manifold \cite{semenov, alekseev:articolo}. The group $D$ equipped with this Poisson bracket  is also called the {Heisenberg double} of $G$. 
 
By writing $\gamma$ as $\gamma= \tilde g g$, with $\tilde g \in \tilde{G}, g\in G$,  it can be shown that these brackets are  compatible with the following ones:
\begin{align} 
 \{\tilde g_1,\tilde g_2\}  &=-[r,\tilde g_1\tilde g_2] \label{tgtg}
\\
\{{\tilde g}_1,g_2\} &=- {\tilde g}_1r  g_2~~~~~\{ { g}_1,{\tilde g}_2\} =- {\tilde g}_2 r^*   g_1  \label{tgg}\\
 \{g_1,g_2\}  &= [r^*,g_1g_2]. \label{gg}
\end{align}
where \eqn{tgtg} and  \eqn{gg} are the  Sklyanin brackets \cite{sklyan, sklyan2}. 
Let us now specify to the example at hand with $G=SU(2)$ and $\tilde{G}=SB(2,\C)$. One can choose for the latter the parametrization $ \tilde g= 2(u_0e ^0+ i u_i \tilde e^i)$ with $u_0^2- u_3^2=1$ and $\tilde e^0=\mathds{1}/2$, $\tilde e^i$ being generators of the Lie algebra $\mathfrak{sb}(2,\C)$, which is going to be specified below.  By expanding   $\tilde g$ as a function of the parameter $\lambda$, $\tilde g(\lambda)= 1+i\lambda I_i e^i + \mathcal{O}(\lambda^2)$, while keeping 
 $g= y^0 \sigma_0 + i y^i \sigma_i$,  one obtains, in the limit $\lambda \rightarrow 0$: 
\beqa
\{I_i,I_j\}&=&\epsilon^k_{\,ij}I_k \nonumber \\  
\{I_i, y^0\}&=& i y^j \delta_{ij} ~~~~~~~~~\{I_i, y^j\}= i y^0 \delta_i^j -  \epsilon_{ik}^j y^k \nonumber \\ 
\{y^0, y^j\}& =& \{y^i, y^j\} = 0 + O(\lambda)\nonumber
\eeqa
which reproduce correctly the canonical Poisson brackets on the cotangent bundle of $SU(2)$. 
Consider now  $r^*$ as an  independent solution of the Yang-Baxter equation  
$r^*\rightarrow \rho= \mu e_k\otimes e^k$ with $ \mu\in\R $ and expand $g\in SU(2)$ as a function of the parameter $\mu$, 
$
g= \mathbf{1} + i \mu \tilde {I}^{i} e_i + O(\mu^2)
$
while keeping $\tilde g$ in its original parametrization. 
By repeating the same analysis  as above, one gets back the canonical Poisson structure on $T^*SB(2,C)$, with position coordinates and momenta now interchanged. In particular we note that:
\be
\{\tilde{I}^i, \tilde{I}^j\}= {f^{ij}}_k \tilde{I}^k \,\, .
 \ee
 Furthermore, it is possible to   consider a different Poisson structure on the double \cite{semenov}, given by: 
\be
\{\gamma_1,\gamma_2 \}= \frac{\lambda}{2}\left[\gamma_1(r^*-r) \gamma_2 - \gamma_2(r^*-r)\gamma_1\right] \,\,\, .
\ee
This is the one that correctly dualizes  the bialgebra structure on $\mathfrak{d}$ when evaluated at the identity of the group $D$.
Indeed, by
  expanding $\gamma\in D$ as $\gamma= \mathbf{1}+ i\lambda I_i \tilde e^i+ i\lambda \tilde{I}^i e_i$  and  rescaling  $r, r^*$  by the same parameter $\lambda$, one can show that:
\beqa
\{I_i, I_j\}&=& {\epsilon_{ij}}^k I_k; ~~~~~~~~~ \{\tilde{I}^i, \tilde{I}^j\}= {f^{ij}}_k \tilde{I}^k  \\
\{I_i, \tilde{I}^j\}&=&  - {f^{ jk}}_i I_k - \tilde{I}^k {\epsilon _{ki}}^{ j}  
\eeqa
which are the Poisson brackets induced by the Lie bi-algebra structure of the double.
One can see that the fiber coordinates $I_i$ and $\tilde{I}^j$  play a symmetric role.
Moreover, since the fiber coordinate $\tilde{I}^i$ appears in the expansion of $g$, it can also be thought of as the fiber coordinate of the {\it tangent} bundle $TSU(2)$, so that the couple $(I_i, \tilde{I}^i)$ identifies the fiber coordinate of the generalized bundle $T\oplus T^*$ over $SU(2)$.

\subsection{The dual model}\label{DIRR}
Let us now go back to the two scalar products in the Lie bialgebra, \eqn{prod1}-\eqn{prod2}. 
With respect to  the second scalar product, one has another splitting:
\be
(e_i, e_j)= - (b_i, b_j)= \delta_{ij}, ~~~~~~~(e_i, b_j)=0 
\end{equation}
with 
maximal isotropic subspaces: ~$f_i^{\pm}= \frac{1}{\sqrt 2}(e_i\pm b_i) $.    
The following  {\it doubled} notation can be introduced:
\be
e_I=\begin{pmatrix}e_i\\ {\tilde e}^i \end{pmatrix},
 \qquad  e_i \in \mathfrak{su}(2), \quad  {\tilde e}^i \in \mathfrak{sb}(2,\mathbb{C}) \,\, . 
\ee

 The first scalar product  then becomes:
\be
\Braket{e_I,e_J}=\eta_{IJ}= 
\begin{pmatrix}
0 & \delta_i^j \\
\delta_j^i &  0
\end{pmatrix}
\ee
which is  $O(3,3)$ invariant  by construction. 

The second scalar product yields:
\be
(e_I,e_J)=
\begin{pmatrix}
\delta_{ij} & \epsilon_{i p 3}\delta^{pj} \\
 \delta^{ip}\epsilon_{j p 3}&  \delta^{ij}- \epsilon^{i k3} \delta_{kl}  \epsilon^{j  l3} \,\,\, 
\end{pmatrix}  \,\,\, .
\ee  
With $C_+, C_-$ being the two  subspaces spanned by $\{e_i\}$, $\{b_i\}$ respectively,  one can  notice that the splitting ${\mathfrak d}= C_{+} \oplus C_{-}$ defines a positive definite metric on ${\mathfrak d}$ via:
\be
\mathcal{H}= (\;,\;)_{C_+}-  (\;,\;)_{C_-}  \,\,\, .   \label{nondegG}
\ee
It is immediate to check that the metric ${\mathcal H}$,  that will be indicated since now on by  double round brackets: 
$$
((e_i,e_j)):= (e_i,e_j); ~~~~~ ((b_i,b_j)):=-(b_i,b_j);~~~~~ ((e_i,b_j)):= (e_i,b_j)=0 
$$
satisfies
$$ {\mathcal H}^T \eta {\mathcal H} = \eta \,\,\, , $$ 
namely ${\mathcal H}$  is a pseudo-orthogonal $O(3,3)$ metric.  The sum  $\alpha \eta+\beta {\mathcal H}$  is a non-degenerate metric for a suitable choice of the parameters $\alpha, \beta$.  Notice that the latter can be rewritten as
\be
((u,v))=2 {\rm Re} \Tr[u^\dag v] \label{noninvprod}
\ee
showing that it is in general not invariant. 

In ref. \cite{MPV18} a dynamical model has been introduced on the cotangent bundle of the dual group $T^*SB(2,\C)$, with action given by:
\be
{\tilde S}_0= -\frac{1}{4} \int_\R {\mathcal Tr}  [\phi^*( {\tilde g}^{-1} \dd {\tilde g}) \wedge \h \phi^*({\tilde g}^{-1} \dd {\tilde g}) ]\label{duac}
\ee
with $\phi: t\in \R\rightarrow \tilde g\in SB(2,\C)$,  $\phi^*$ the pull-back map, $\tg= 2( u_0 \te^0 
+ i  u_i \te^i)$, and $u_0^2- u_3^2=1$.     ${\mathcal Tr} $ was chosen to be  the non-degenerate product \eqn{nondegG}  {${\mathcal Tr} := ((\; , \;))$}, which is however only invariant under left $SB(2,\C)$ action \cite{MPV18}. The latter defines a non-degenerate left-invariant metric over the fibers,
\be
h^{ij}:= \delta^{ij } +\epsilon^{ik3}\delta_{kl}\epsilon^{jl3} \label{hij}
\ee
so that the Lagrangian can be rewritten as:
 \be
\tilde { {L}}_0=\frac{1}{2} {\dot {\tilde Q}}_i h^{ij}{\dot{ \tilde Q}}_j 
\ee   
with ${\dot{ \tilde Q}}_i =  u_0 \dot u_i - u_i \dot u_0 +{ f^{jk}}_i u_j \dot u_k$  being the left tangent bundle coordinates defined through the Maurer-Cartan form:  
\be {\tilde g}^{-1}  \dot {\tilde g}= 
{\dot{ \tilde Q}}_i\tilde e^i.
\ee
In analogy with the case of the rigid rotor, the equations of motion are easily retrieved:
\be
{\sf L}_{\Gamma} \dot {\tilde Q}_j h^{ji} -  \dot {\tilde Q}_p \dot {\tilde Q}_q {f^{ip}_k} h^{qk} = 0
\ee
where  ${\sf L}_\Gamma$ is the Lie derivative with respect to $\Gamma=\frac{d}{dt}$. 
   
Left  momenta living on the cotangent bundle are introduced through a Legendre transform:
\be
{\tilde I}^i= \frac{\del \tilde {\mathcal L}_0}{\del {\dot{\tilde Q}}_i}= h^{ij} {\dot{\tilde  Q}}_j
%= \textcolor{blue}{-\frac{i}{2}(({\tilde g}^{-1}\dot{\tilde g},{\tilde e}^j))}
\ee
%with inverse relation 
  %  ${\dot{\tilde  Q}}_j= (\delta_{jr} -\frac{1}{2}\epsilon_{jr3}) {\tilde I}^r $. The Hamiltonian reads then 
\be
\tilde{ {H}}_0= \frac{1}{2}{\tilde I}^i h_{ij} {\tilde I}^j 
\ee
with
\be \label{invh}
h_{ij}=(\delta_{ij} -\frac{1}{2} \epsilon_{ik3}\delta^{kl}\epsilon_{jl3} )
\ee
the inverse metric.  By means of the 
Poisson brackets (see ref. \cite{MPV18} for details):
\beqa
\{u_i, u_j\}&=&0 \label{uu}\\
\{ {\tilde I}^i, {\tilde I}^j \}&=& {f^{ij}}_k {\tilde I}^k \label{tItI}\\
\{u_i, {\tilde I}^j \}&=& \delta_i^j u_0 - {f^{jk}}_i u_k \label{utI}
\eeqa
one obtains the Hamiltonian dynamics 
\be
\dot{\tilde I}^j = {f^{jk}}_l {\tilde I}^l {\tilde I}^r {h}_{rk}
\ee
expressing that the Hamiltonian is not invariant under right $SB(2,\C)$ action. By introducing right momenta we would get instead $\dot{\underline I}^j=0$, consistently with the invariance of the Hamiltonian under left action. 
    
The Poisson brackets of both models, reported in Eqs. \eqn{yy}-\eqn{yI} and \eqn{uu}\eqn{utI}, have the structure of a semi-direct product. Moreover  the Poisson brackets for the momenta can be retrieved by the Poisson-Lie bracket of the dual group (resp. Eq. \eqn{tgtg} for the Poisson bracket of the $SU(2)$ momenta, Eq. \eqn{gg} for the Poisson bracket of the $SB(2,\C)$ momenta). It is therefore natural to describe this structure as a kind of Poisson-Lie duality and look for a generalized model over the group manifold of the double group, which encodes both models once suitably constrained. 
   
\subsection{The generalized action}\label{genacsec}
In ref. \cite{MPV18} a generalized action with doubled degrees of freedom has been introduced in the form:
\be\label{genacIRR}
\mathcal{S}= \int k_1 \Braket{\gamma^{-1} \dd\gamma \wedge \h \gamma^{-1} \dd\gamma} +k_2 ((\gamma^{-1} \dd\gamma\wedge \h \gamma^{-1} \dd\gamma))
\ee
with  ~~ $\gamma\in SL(2,\C)$, ~~ $e_I=(e_i, \tilde e^i)$, 
$ \gamma^{-1} \dd \gamma= \dot {\bf Q}^I e_I dt  \equiv  (A^i e_i + B_i \tilde e^i) dt
$ the left-invariant Maurer-Cartan one-form on $SL(2,\C)$ pulled-back to $\R$ and 
$(A^i, B_i)$  fiber coordinates of $TSL(2,\C)$ .   
 They  are obtained by means of the  scalar product  \eqn{prod1} according to:
\be\label{AiBi}
A^i= 2{\rm Im} \Tr ( \gamma^{-1}\dot \gamma \tilde e^i); \;\;\; B_i= 2{\rm Im} \Tr (\gamma^{-1}\dot \gamma  e_i).
\ee
Upon introducing $k=k_1/k_2$, the Lagrangian can be rewritten in terms of the left generalized coordinates
$\dot {\bf Q}^I$ as follows:
\be
{L}=   \frac{1}{2}  (k ~\eta_{IJ}+ \mathcal{H}_{IJ} ) \dot {\bf Q}^I  \dot {\bf Q}^J    
 \ee    
 with 
\be
k \eta_{IJ}+ \mathcal{H}_{IJ}= 
\begin{pmatrix}
 \delta_{ij} &(k \delta_{i p}+  \epsilon_{i p 3}) \delta^{pj} \\
 \delta^{i p }(k \delta_{p j}-\epsilon_{j  p 3} ) & \delta^{ij}+ \epsilon^{i k3} \delta_{kl} \epsilon^{j l3}
\end{pmatrix}   \,\,\, .
\ee   
The equations of motion are:
\be
{\sf L}_\Gamma\dot {\bf Q}^I( k\, {\cal \eta}_{IJ}+ \, {\cal H}_{IJ}) -\dot {\bf Q}^P \dot {\bf Q}^Q {C_{IP}}^K ( k\, {\cal \eta}_{QK}+ \,{\cal H}_{QK})  =0 \label{eomd}
\ee
where ${C_{IP}}^K$ are the structure constants of $\mathfrak{sl}(2,\C)$. The  matrix   $k\eta_{IJ}+ \mathcal{H}_{IJ}$ is non-singular provided $k^2 \ne 1$, which is going to be assumed from now on.     
In the doubled description introduced above, the left generalized momenta are represented by:
\be
{\bf P}_I = \frac{\del  L}{\del \dot {\bf Q}^I}= ( k {\cal \eta}_{IJ}+ \,{\cal H}_{IJ})\dot {\bf Q}^J \,\,\, .  \label{genP}
\ee
The Hamiltonian reads then as:
\be
{\hat H}= ({\bf P}_I \dot{\bf Q}^I -  L)_{{\bf P}}= \frac{1}{2} [( {k \cal \eta}+  \, {\cal H})^{-1}]^{IJ} {\bf P}_I{\bf P}_J   \ee
with 
\be
[( {k\cal \eta} +  {\cal H})^{-1}]^{IJ}= \frac{1}{1-k^2} 
\begin{pmatrix}
\delta^{ij}+ \epsilon^{i l3}\delta_{lk} \epsilon^{j k3}& -(\epsilon^{i p 3}+ k \delta^{i p}) \delta_{p j}
\\
(\epsilon_{i p 3}-k \delta_{i p}) \delta^{p j}& \delta_{ij}  \,\,\, 
\end{pmatrix}  \,\, .    \nonumber
\ee
In terms of the components  $I_i, \tI^j$ of ${\bf P}_I$  the Hamiltonian can be rewritten as:
%From  \eqn{genP} one can explicitly write the generalized momenta ${\bf P}_I$ in terms of the components of $\dot{\bf Q}^I\equiv(A^i, B_j)$, finding:
%\be
%{\bf P}_I \equiv ( I_i, \tI^i)=\left(\delta_{ij} A^j+(k\delta_i^j+ \epsilon_i^{j3})B_j, (k \delta^i_j-\epsilon^i_{j3})A^j+[\delta^{ij}+\delta^{lk}\epsilon^i_{l3}\epsilon^j_{k3}]B_j\right).  \nonumber
%\ee
\be \label{Hdoubirr}
{\hat H} =\frac{1}{2( 1-k^2)}\left( ( \delta^{ij}  + \epsilon^{i l3}\delta_{lk} \epsilon^{j k3})  I_i I_j+  \delta_{ij} \tI^i \tI^j  - 2(\epsilon^{i p 3}+ k \delta^{i p}) \delta_{p j} I_i \tI^j \right) 
\ee
% \\
%&=& 
%\frac{1}{2}(1-k^2)^{-1} \left ( (1-k^2) \delta^{ij} I_i I_j + \delta_{ij}(\tI^i -I_s(k \delta^{si}+ {\epsilon^{si}}_3))
%(\tI^j -I_r(k  \delta^{rj}+ {\epsilon^{rj}}_3))\right)\nn
%\eeqa
%which can be rewritten as
%\be
%{H} =\frac{1}{2}(1-k ^2)^{-1}\left((1-k ^2) \delta^{ij} I_i I_j + \delta_{ij}\tilde{\cal I}^i \tilde{\cal I}^j\right)  
%\ee
%after having defined
%\be
%\tilde{\cal I}^i \equiv \tI^i -I_s(k  \delta^{si}+ {\epsilon^{si}}_3)= \delta^{ij} (1-k^2) B_j.  \nonumber
%\ee
with Poisson brackets (see ref. \cite{MPV18} for a derivation) 
%In order to obtain the Hamilton equations for the generalized model on the Drinfel'd double,  one can proceed as in the previous section with the determination of Poisson brackets from the first-order action functional:
%\be
%{\widehat{\mathcal{S}}}= \int \langle {\bf P} |  \gamma^{-1}d \gamma\rangle  - \int \widehat{H} dt \equiv \int  {\boldsymbol{\theta}} -\int \widehat{H} dt  \nonumber
 %\ee
%with 
%\beqa
%{\bf P}&=& i\; {\bf P}_I {e^I}^*= i\;(I_i{e^i}^* + \tI_i \te_i^*) \nonumber \\
%\gamma^{-1}d\gamma&=& i\,  {\boldsymbol{\alpha}}^J e_J= (\alpha^k e_k + \beta_k \te^k)  \nonumber  \,\, .
%\eeqa
%We stress once again that ${\bf P_I}$, $ {\boldsymbol{\alpha}}^J$ are respectively generalized momenta and basis one-forms on the doubled configuration space $SL(2,\C)$. The symplectic form on $T^*SL(2,\C)\simeq SL(2,\C)\times \mathfrak{sl}(2,\C)^*$ is therefore:
%\beqa
%{\boldsymbol{\omega}}= d {\boldsymbol{\theta}}&=& dI_i\wedge \alpha^i + d\tI^i\wedge  \beta_i +\frac{1}{2}\tI^l\left(\alpha^j\wedge\beta_k {\epsilon^k}_{jl}- \beta_j\wedge \alpha^k {\epsilon^j}_{kl}- \beta_j\wedge\beta_k {f^{jk}}_l\right)\nn\\
%&+&  \frac{1}{2}I_l\left(-\alpha^j\wedge\alpha^k {\epsilon^l}_{jk}+ \alpha^j\wedge \beta_k {f^{lk}}_{j}- \beta_j\wedge\alpha^k {f^{lj}}_k\right)  \nonumber
%\eeqa
%which  yields for the generalized momenta the Poisson brackets:
\beqa \label{remark}
\{I_i, I_j\}&=& {\epsilon_{ij}}^k I_k \label{remark1}\\
\{\tI^i, \tI^j\}&=& {f^{ij}}_k \tI^k\\
\{I_i, \tI^j\}&=& {\epsilon_{il}}^j \tI^l- I_l {f^{lj}}_i  \;\;\;\;\{\tI^i, I_j\}= -{\epsilon_{jl}}^i \tI^l+ I_l {f^{li}}_j \label{remark3}
\eeqa
while the Poisson brackets between momenta and configuration space variables $g,\tg$ are unchanged with respect to $T^*SU(2), T^*SB(2,\C)$. 
%We shall come back to the Poisson algebra \eqn{remark} in a while. 

In order to derive Hamilton equations, it is sufficient to write in compact form:
\be
\{{\bf P}_I, {\bf P}_J\}= {C_{IJ}}^K {\bf P}_K   \nonumber
\ee
with ${C_{IJ}}^K$ the $SL(2,\C)$ structure constants as specified above in Eqs. \eqn{remark1}-\eqn{remark3}. We have then: 
\be
\frac{d}{dt} {\bf P}_I= \{ {\bf P}_I, \widehat H\}=  [( {\cal \eta}+ k\, {\cal H})^{-1}]^{JK} \{ {\bf P}_I, {\bf P}_J\} {\bf P}_K= [( {\cal \eta}+ k\, {\cal H})^{-1}]^{JK} {C_{IJ}}^L {\bf P}_L{\bf P}_K   \nonumber
\ee
which is not zero, consistently with \eqn{eomd}.

 Summarizing, 
 \begin{itemize}
 \item we have obtained a dynamical model with doubled coordinates and generalized momenta;
 \item the Hamiltonian dynamics is dictated by   Poisson brackets for the generalized momenta which reproduce the bialgebra structure of $\mathfrak{sl}(2,\C)$.
 \end{itemize}
  These brackets can be obtained  from the following  Poisson structure on the double, first introduced in ref. \cite{semenov} : 
\be
\{\gamma_1,\gamma_2 \}= \frac{\lambda}{2}\left[\gamma_1(r^*-r) \gamma_2 - \gamma_2(r^*-r)\gamma_1\right].\label{gammargamma}
\ee
This is the one that correctly dualizes  the bialgebra structure on $\mathfrak{d}$ when evaluated at the identity of the group $D$. To this, let us expand $\gamma\in D$ as $\gamma= \mathds{1}+ i\lambda I_i \tilde e^i+ i\lambda \tI^i e_i$  and rescale  $r, r^*$  by the same parameter $\lambda$. 
It is straightforward to obtain, on the l.h.s. of Eq. \eqn{gammargamma}, the following expression:
\be
\{\gamma_1,\gamma_2 \}= -\lambda^2 \left(\{I_i, I_j\} \tilde e^i\otimes \tilde e^j + \{\tI^i, \tI^j\}  e_i\otimes  e_j  +\{I_i, \tI^j\} (\tilde e^i\otimes  e_j- e_j \otimes \tilde e^i)   \right)  \nonumber
\ee
while, on the r.h.s. of the same equation, one gets:
\be
-\lambda^2 \left(I_s \epsilon^s_{\,ij} \tilde e^i\otimes  \tilde e^j+ \tI^s f_s^{\,ij}  e_i\otimes   e_j+ I_s f_i^{\,sj}(\tilde e^i\otimes e_j-e_j\otimes \tilde e^i ) + \tI^s \epsilon^j_{\,si}(\tilde e^r\otimes  e_j -  e_j\otimes \tilde e^i)\right) \,\,.  \nonumber
\ee
By equating the two results, one reproduces the Poisson algebra \eqn{remark1}-\eqn{remark3}, which is the wanted result. 
%\beqa
%\{I_i, I_j\}&=& {\epsilon_{ij}}^k I_k  \nonumber \\
%\{\tI^i, \tI^j\}&=& {f^{ij}}_k \tI^k  \nonumber \\
%\{I_i, \tI^j\}&=&  - {f_i}^{ jk}I_k - \tI^k {\epsilon _{ki}}^{ j}  \nonumber
%\eeqa
%which is nothing but the Poisson bracket induced by the Lie algebra structure of the double \eqref{liemis}. 

Upon using the  compact notation $I= i I_i {e^i}^*, \tI = i \tI^i {\te_i}^*$, with ${e^i}^*$, ${\te_i}^*$ respectively representing the dual bases of  ${e_i}$, ${\te^i}$, one can rewrite the Poisson algebra as follows: 
\be
\{I +\tI, J+\tilde{J}\}= \{I,J\} -\{J, \tI\}+ \{I,\tilde{J}\} + \{\tI,\tilde{J}\}. \label{cb}
\ee
which is argued in ref. \cite{MPV18} to  represent a Poisson realization of a C-bracket for the generalized bundle $T\oplus T^*$ over $SU(2)$. We refer to ref. \cite{MPV18} for details. 
%, once  considering the isomorphisms 
%\be
%TSL(2,\C)\simeq SL(2,\C)\times \mathfrak{sl}(2,\C)  \nonumber
%\ee
%with the fiber:
%\be
 %\mathfrak{sl}(2,\C)\simeq \mathfrak{su}(2)\oplus \mathfrak{sb}(2,\C)\simeq TSU(2)\oplus T^*SU(2).  \nonumber
 %\ee
%That is, we recognize $I= i I_i {e^i}^*, J= i J_i {e^i}^*$ as one-forms, with ${e^i}^*$ being a basis over $T^*$ and  $\tI = \tI^i {\te_i}^*, \tilde J= \tilde J^i {\te_i}^*$ as  vector-fields, with ${\te_i}^*$ a basis over $T$. Namely, the couple $(I_i, \tI^i)$ identifies the fiber coordinate of the generalized bundle $T\oplus T^*$ of $SU(2)$

In order to complete the analysis, let us look at the Lie algebra of Hamiltonian vector fields associated with the momenta $I, \tI$.
Hamiltonian vector fields are  defined in terms of Poisson brackets in the standard way:
\be
X_f \equiv \{\cdot \;, f\}  \label{hamvecfi}
\ee
so that, by indicating with  $X_i= \{\cdot \;, I_i\}, \tilde X^i= \{\cdot \;, \tilde I^i\}$ the Hamiltonian vector field associated with $I_i, \tilde I^i$ respectively, one has, after using the Jacobi identity,  the following Lie algebra:
\beqa\label{havefialgebra}
[X_i, X_j] &=& \{\{\cdot \;, I_j\}, I_i\}-\{\{\cdot \;, I_i\}, I_j\}=\{\cdot \; ,\{I_i,I_j\}\}={ \epsilon_{ij}}^k\{\cdot \; ,I_k\}= {\epsilon_{ij}}^kX_k \label{havefialgebra1} \\
{[}{\tilde X}^i, {\tilde X}^j{] }&=& \{\{\cdot \;, \tI^j\}, \tI^i\}-\{\{\cdot \;, \tI^i\}, \tI^i\}=\{\cdot \; ,\{\tI^i,\tI^j \} \}={ f^{ij}}_k\{\cdot \; ,\tI^k\}= {f^{ij}}_k\tX^k  \label{havefialgebra2}\\
{[}X_i,{\tilde X}^j{] }&=& \{\{\cdot \;,  \tI^j \},I_i\}-\{\{\cdot \;, I_i\}, \tI^j\}=\{\cdot \; ,\{I_i,\tI^j\}\}=  - {f_i}^{ jk}\{\cdot \;,I_k\} - \{\cdot \;, \tI^k\} {\epsilon _{ki}}^{ j} \nonumber\\
&=& - {f_i}^{ jk} X_k -{\tilde X}^k  {\epsilon _{ki}}^{ j}  \label{havefialgebra3}
\eeqa
namely:
\be
[X+\tX, Y+\tY]= [X,Y]+ {\sf L}_X \tY-{\sf L}_Y \tX + [\tX, \tY]  \nonumber
\ee
which  shows that  C-brackets can be obtained as  derived brackets, in analogy with  the ideas of  ref.s \cite{deser1, deser2}, with the remarkable difference that, in this case, they are  derived from the canonical Poisson brackets of the dynamics. 

 In order to get back one of the two models with half degrees of freedom one has to impose constraints. This has been realized in ref. \cite{MPV18} by gauging the global symmetries of the generalized action, namely the  $SU(2)$ or $SB(2,\C)$ global invariance. The same procedure will be adopted for the chiral model below, therefore we refer again to ref.  \cite{MPV18} for details about the gauging of the generalized model described above.

\section{The Principal Chiral Model}\label{PCM}
A Principal Chiral Model is a two-dimensional field theory with target configuration space given by a Lie group $G$ and source space given by the two-dimensional spacetime $\mathbb{R}^{1,1}$ endowed with the metric $s_{\alpha \beta}=\mathrm{diag}(1,-1)$. 

The $SU(2)$ Principal Chiral Model represents a natural generalization to field theory of the dynamics of the IRR, as described above. Indeed, the action functional is formally the same, while the field variables are defined on two-dimensional spacetime taking values on the group manifold of $SU(2)$. The possibility of introducing a one-parameter family of Hamiltonian descriptions with modified Poisson brackets,  yielding the same equations of motion, was already illustrated  in ref.s \cite{R89}, \cite{RSV93}, \cite{RSV96}. We are going to follow that approach in order to show that it naturally yields a family of dually related models. The duality transformations which we shall find will be shown to be of Poisson-Lie type.

% and \cite{edwards:nonpub}.
%Let us recall that an $O(n)$ non linear sigma model is a field theory with target space given by a sphere $S^n$ and a two-dimensional manifold as base space.\\ 

In the Lagrangian approach  the action may be written in terms of fields $\phi : (t,\sigma) \in  \R^{1,1}\rightarrow g\in  SU(2)$ and Lie algebra valued  left-invariant one-forms whose pull-back to $\R^{1,1}$ may be written as 
 \be
  \phi^*(g^{-1}\mathrm{d}g)=(g^{-1}\partial_{t}g )\,\mathrm{d}t+(g^{-1}\partial_{\sigma}g) \,\mathrm{d}\sigma
  \ee
%  with $e_i$ the Lie algebra generators of $SU(2)$, 
so to have: 
%\be\label{startingac}
%S=\textcolor{red}{\frac{1}{2}}\int_{\mathbb{R}^2}\Tr[\phi^*(g^{-1}\mathrm{d}g) \wedge \h \phi^*(g^{-1}\mathrm{d}g)],  
%\ee
\be\label{startingac}
 S= \frac{1}{4} \int_{\mathbb{R}^2}\Tr[\phi^*(g^{-1}\mathrm{d}g) \wedge \h \phi^*(g^{-1}\mathrm{d}g)]
\ee

where  trace is understood as the scalar product in the Lie algebra $\mathfrak{su}(2),$ and the Hodge star operator acting as $\h \dd t=  \dd\sigma, \h\dd\sigma= \dd t$ \footnote{We adopt the the convention $\epsilon_{01}=1$.}, yielding:
%\be
%S= \textcolor{red}{\frac{1}{2}}\int_{\mathbb{R}^2}\mathrm{d}t\mathrm{d}\sigma\ \Tr\bigl[ (g^{-1}\partial_tg)^2 - (g^{-1}\partial_{\sigma}g)^2 \bigr] \label{act0} 
%\ee
\be
S=  \frac{1}{4}\int_{\mathbb{R}^2}\mathrm{d}t\mathrm{d}\sigma\ \Tr\bigl[ \{ (g^{-1}\partial_t g)^2   - (g^{-1}\partial_{\sigma}g)^{2}  \bigr]  \label{act1}
\ee
%\textcolor{blue}{
%\be
%S=  \frac{1}{4}\int_{\mathbb{R}^2}\mathrm{d}t\mathrm{d}\sigma\ \Tr\bigl[ \{ (g^{-1}\partial_tg)^{i} (g^{-1}\partial_tg)^{j}  - (g^{-1}\partial_{\sigma}g)^{i} \partial_{\sigma}g)^{j}) \}e_{i} e_{j} \bigr]
%\ee
%}
which is to be compared with \eqn{lag} for the IRR dynamics.
%The analogy with   the IRR  dynamics is obtained by choosing the Lie group  $SU(2)$ as target space for the dynamics , i.e. $\phi: \mathbb{R}^{1,1} \rightarrow SU(2)$, so that the action $S$ is written in terms of  the  left-invariant one-form $g^{-1}\mathrm{d}g$ that takes values in $\mathfrak{su}(2)$, in which case   the scalar product  in Eq. \eqn{act0} is effectively  the trace over the Lie algebra. 
%Being $ \Tr\bigl( e_i e_j \big) = \frac{1}{2} \delta_{ij}$, one thus has:
%\be
%S= \textcolor{red}{\frac{1}{4}}\int_{\mathbb{R}^2}\mathrm{d}t\mathrm{d}\sigma\ \bigl[  (g^{-1}\partial_tg)^i   (g^{-1}\partial_tg)^j -  (g^{-1}\partial_{\sigma}g)^i  (g^{-1}\partial_{\sigma}g)^j\ \bigr]\delta_{ij}. \label{act1}
%\ee
%\textcolor{blue}{
%\be
%S= \frac{1}{8}\int_{\mathbb{R}^2}\mathrm{d}t\mathrm{d}\sigma\ \bigl[  (g^{-1}\partial_tg)^i   (g^{-1}\partial_tg)^j -  (g^{-1}\partial_{\sigma}g)^i  (g^{-1}\partial_{\sigma}g)^j\ \bigr]\delta_{ij}.
%\ee
%}
%it being
%\be
%g^{-1}\partial_t g=(g^{-1}\partial_t g)^ie_i, \quad g^{-1}\partial_{\sigma}g=(g^{-1}\partial_{\sigma}g)^ie_i,
%\ee
A remarkable property of the model is that its  Euler-Lagrange equations
\begin{equation}
\partial_t(g^{-1}\partial_t g)-\partial_{\sigma}(g^{-1}\partial_{\sigma}g)=0  
\end{equation}
may be rewritten in terms of  an equivalent system of two first order partial differential equations, introducing the so called  \emph{currents}, as it is customary in the framework of integrable systems:
\begin{equation}
A^i=\Tr (g^{-1}\partial_t g) e_i, \quad {J^i}=\Tr ( g^{-1}\partial_{\sigma}g) e_i, \label{curr}
\end{equation}
%\textcolor{blue}{
%\begin{equation}
%A= (g^{-1}\partial_t g )^{i} e_i, \quad {J}=(g^{-1}\partial_{\sigma}g)^{i} e_i, \label{curr}
%\end{equation}
%}
namely, $g^{-1}\del_t g = 2 A^i e_i, g^{-1}\del_\sigma g = 2 J^i e_i$, with $\Tr e_i e_j=\frac{1}{2} \delta_{ij}$.  The Lagrangian  becomes:
\be
L= \frac{1}{2}\int_\R \dd\sigma (A^i\delta_{ij} A^j - J^i \delta_{ij} J^j) \label{unmodlag}
\ee
with
\begin{align} 
\partial_t A= & \partial_{\sigma} J,  \label{equiv1} \\
\partial_t J= & \partial_{\sigma} A-[A,J].  \label{equiv2}
\end{align}
The existence of a $g \in SU(2)$ that admits the expression of the currents in the form \eqref{curr} is guaranteed by Eq.  \eqref{equiv2}, that can be read as an integrability condition. Moreover, if  the usual boundary condition for a physical field is imposed:
\be
 \lim_{\sigma\to\pm\infty} g(\sigma)=1, \label{boundco} 
 \ee
 one has that $g$ is uniquely determined from \eqref{curr}.\footnote{
Note \cite{R89} that if we had chosen space to be a circle, \eqn{equiv2} would not imply \eqn{curr}. The solution to these equations will not be periodic in general. If $(A, J)$ is viewed as a connection, \eqn{equiv2} says that it is flat. But in order for a  flat connection to be `pure gauge' as in \eqn{curr}, it is necessary also for the parallel transport operator around a homotopically non-trivial curve (holonomy) to be equal to the identity. }
At fixed $t$, all the elements $g$ satisfying the boundary condition \eqn{boundco}  form an infinite dimensional Lie group $SU(2)(\mathbb{R})\equiv \mathrm{Map}(\R,SU(2))$, given by smooth maps $g: \sigma\in \mathbb{R}\rightarrow g(\sigma)\in SU(2)$ which are constant at infinity \cite{R89}. This is a slight generalization of the definition of  loop group which is the group of smooth maps from $S^1$ to $SU(2)$. 

 At fixed time, the currents $J$ and $A$ take values in the Lie algebra $\mathfrak{su}(2)(\mathbb{R})$, defined as the algebra of functions from $\mathbb{R}$ to $\mathfrak{su}(2)$ that are sufficiently fast decreasing at infinity to be square-integrable. Again, this definition generalizes the one  of  loop algebra $\mathfrak{g}({S^1})$, which,  for $\mathfrak{g}$  a semi-simple Lie algebra, are known as  Kac-Moody algebras.  
 
 The analogy with particle dynamics on Lie groups can be pushed further, by regarding  the carrier space of the dynamics as the tangent bundle  of $SU(2)(\mathbb{R})$.  Therefore the tangent bundle description of the dynamics can be given in terms of $(J,A)$, with $A$ being the left generalized velocities and $J$ playing the role of left configuration space coordinates.

%We choose these functions to be from $\mathbb{R}$ in order to have equal-time (Poisson) Lie bracket on  $\mathfrak{g}(\mathbb{R})$, as usual in field theory. Substantially, we can write the elements in this Lie algebra as $I= I^i (\sigma)e_i$ and $J=J^i(\sigma)e_i$.\\
%Let us state this in a proper form \cite{sinistro}.
%\theoremstyle{definition}
%\begin{definition}
%Let $G$ be a compact Lie group. The group $G(M)$, or $Map(M,G)$, associated with $G$ over a smooth manifold $M$ is the Lie group given by the smooth mappings $$g: M \ni x \rightarrow g(x) \in G,$$
%where the pointwise group law is the composition of functions $$g''(x)=g'(x)\circ g(x), \ \mathrm{with}\  g''(x),g'(x),g(x) \in G(M),$$ the inverse element is $g^{-1}=[g(x)]^{-1}$ and the unit element is the identity in $G$, $e(x)=e$. \\
%The group $G(M)$ is infinite-dimensional.
%\end{definition}
Infinitesimal generators of the Lie algebra $\mathfrak{su}(2)(\R)$ can be obtained by considering the vector fields which generate the  finite-dimensional Lie algebra $\mathfrak{su}(2)$ and replacing  ordinary derivatives  with functional derivatives, thus yielding 
\begin{equation}
X_i(\sigma)=X_i^a(\sigma) \frac{\delta \ \ \ }{\delta g^a(\sigma)},   \label{funder}
\end{equation}
and their Lie bracket is 
\begin{equation}
[X_i(\sigma),X_j(\sigma')]=c_{ij}^{\ \ k} X_k (\sigma) \delta  (\sigma-\sigma'),  
\end{equation}
where $\sigma,\sigma' \in \R$.  This Lie bracket is $C^{\infty}(\R)$-linear and 
$\mathfrak{su}(2)(\R) \simeq \mathfrak{su}(2) \otimes C^{\infty}(\R)$. \\
Notice that the real line $\R$ can be replaced by any smooth manifold $M$.  The Lie algebras $\mathfrak{g}(M)=\mathrm{Map}(M,\mathfrak{g})$ are the so called  \emph{current algebras}.

\subsection{The  Hamiltonian Formulation} \label{hamformul}
Let us briefly review  the standard Hamiltonian approach which can be found for example in \cite{Witten1983, RB1984}. Having recalled in previous section that the target space where the Lagrangian dynamics takes place is  the tangent bundle $TSU(2)$, we shall see in present section  that in the  Hamiltonian  framework the target phase space is naturally given  by $T^*SU(2)$. 
In order to introduce the   canonical formalism,  the  canonical momenta are defined as:
\begin{equation}
I_i=\frac{\delta L}{\delta \ (g^{-1}\partial_t g)^i}=\delta_{ij}(g^{-1}\partial_tg)^j= \delta_{ij} A^j \,\,\, .
\end{equation}
Thus, the Hamiltonian can be written as:
\be
H=\frac{1}{2}\int_{\mathbb{R}} \mathrm{d}\sigma (I_iI_j \delta^{ij} +J^i J^j \delta_{ij}),\label{undefoH}
\ee
 while the equal-time 
Poisson brackets \cite{Witten1983, RB1984} can be checked  to be (see appendix \ref{appA} for a pedagogical derivation)
%
%and the equations of motion \eqref{equiv1} and \eqref{equiv2} can be recovered considering a particular form for the Poisson brackets:
\begin{align} 
\{I_i(\sigma),I_j(\sigma')\}= &{ \epsilon_{ij\;}}^k I_k(\sigma)\delta(\sigma-\sigma'),   \label{IIbr}\\
\{I_i(\sigma),J^j(\sigma')\}= & {\epsilon_{ki\;}}^j J^k(\sigma) \delta(\sigma-\sigma')-\delta_{i}^j\delta'(\sigma-\sigma'), \label{mixbrac} \\
\{J^i(\sigma),J^j(\sigma')\}= & 0,  \label{JJbr}
\end{align}
%induced by the Lie bracket defining the algebra $\mathfrak{su}(2)$, because the canonical momentum $I^i$ is defined on the fiber $\mathbb{R}^3$ of $T^*SU(2)$ \footnote{Following \cite{RSV93} and \cite{RSV96} we can show, with the Sugawara construction, that the Principal Chiral Models are classical conformal invariant field theories.}. In the second bracket, the term proportional to the derivative of the delta function corresponds to a central extension of the algebra.\\
yielding the equations of motion for the momenta:
\begin{equation}
%\begin{split}
%&
 \partial_t I_j(\sigma)=  \{H,I_j (\sigma)\}=
%\frac{1}{2} \int_{\mathbb{R}} \mathrm{d}\sigma \ \bigl(\{I_i(\sigma),I_j(\sigma')\} I_k(\sigma)+I_i(\sigma)\{I_k(\sigma),I_j(\sigma')\}+ \\
%& \qquad \qquad+ \left(\{J^p(\sigma),I_j(\sigma')\} J^q(\sigma)+J^p(\sigma)\{J^q(\sigma),I_j(\sigma')\} \right)\delta_{pi}\delta_{qk}\bigr)\delta^{ik}=\\ 
%&=  \int_{\mathbb{R}}\mathrm{d}\sigma \ \bigl[{\epsilon_{ij}}^k I_k(\sigma)\delta(\sigma-\sigma') I_i(\sigma) +  \bigl(\epsilon_{ij k} J^k(\sigma) \delta(\sigma-\sigma')-\delta_{ij}\delta'(\sigma-\sigma')\bigr)J^i(\sigma) \bigr]= \\
%&=
  \partial_{\sigma}J^k \delta_{kj}(\sigma),  \label{eomI}
%\end{split}
\end{equation}
where we have used the antisymmetry of the structure constants and the integration by parts.  In a similar way, we get the remaining  equations:
\begin{equation}
\partial_t J^j(\sigma)=\{H,J^j(\sigma)\}=\partial_{\sigma}I_k \delta^{kj}(\sigma) -{\epsilon^{\ jl}}_{k}I_lJ^k(\sigma). \label{eomJ}
\end{equation}
The brackets \eqn{IIbr}-\eqn{JJbr} show that $I$ and $J$ span the infinite-dimensional  current algebra $\mathfrak{c}_1$. In particular, the $I$'s are the generators of the affine Lie algebra $\mathfrak{su}(2)(\mathbb{R})$, while the $J$'s span an Abelian algebra $\mathfrak{a}(\R)$, so that $\mathfrak{c}_1$ is the semi-direct sum  $\mathfrak{c}_1=\mathfrak{su}(2)(\mathbb{R})\ltimes \mathfrak{a}(\R)$.  

%Let us notice  here that, because of its structure of semidirect  sum, the algebra  \eqn{IIbr}-\eqn{JJbr} has been considered by many authors in relation with deformations, as a result of a Lie algebra contraction from some semisimple affine algebra. Notab

As noticed before, if one extends the analogy  with the Lagrangian description of particle dynamics on Lie groups to the Hamiltonian setting, the target phase space of the dynamics can be recognized to be the cotangent bundle of $SU(2)$, with   the currents $(J^i, I_i)$ playing the role of conjugate variables and $I$ the left generalized momenta, while $J$ keeping the role of left configuration space coordinates.

A remarkable result due to Rajeev \cite{R89,R892} consists in  the fact that  an equivalent description of the dynamics can be given,  in terms of a new, one-parameter family, of Poisson algebras {\it and}  modified Hamiltonians. Upon introducing a  parameter $\tau$, real or imaginary, the deformed  brackets read as:
\begin{align} 
\{I_i(\sigma),I_j(\sigma')\}= & (1-\tau^2){\epsilon_{ij}}^k I_k(\sigma)\delta(\sigma-\sigma'),   \label{modipo1}\\
\{I_i(\sigma),J^j(\sigma')\}= & (1-\tau^2)J^k(\sigma){ \epsilon_{ki}}^j  \delta(\sigma-\sigma')-(1-\tau^2)^2\delta_{i}^j\delta'(\sigma-\sigma'),  \label{modipo2} \\
\{J^i(\sigma),J^j(\sigma')\}= & (1-\tau^2)\tau^2 {\epsilon^{ij}}_k I_k(\sigma)\delta(\sigma-\sigma'). \label{modipo3}
\end{align}
The modified Hamiltonian reads in turn as:
\begin{equation}
H_{\tau}=\frac{1}{2(1-\tau^2)^2}\int_{\mathbb{R}} \mathrm{d}\sigma \  (I_iI_j \delta^{ij}+J^i J^j \delta_{ij}). \label{modiha}
\end{equation}
and, in the limit $\tau \rightarrow 0$, the algebra and the Hamiltonian reduce to the original ones.  Notice that the factor ($1-\tau^2)$ is never zero for imaginary $\tau$. \\
The new brackets correspond to the infinite-dimensional Lie  algebra $\mathfrak{c}_2$ which, for imaginary $\tau$, our choice from now on,  can be easily recognized to be isomorphic to the current algebra modeled on the Lorentz algebra $\mathfrak{sl}(2,\C)$, that is  $\mathfrak{c}_2\simeq\mathfrak{sl}(2,\mathbb{C})(\R)$\footnote{For real $\tau$ it is instead isomorphic to the algebra $\mathfrak{so}(4)(\R)$.  The latter   case is the one analyzed in detail in  \cite{R89,R892,RSV93,RSV96} with respect to quantization and integrability. Here we stick  to imaginary $\tau$, this being the choice which unveils  the double group structure. }.  The Lie algebra $\mathfrak{c}_1$ can be recovered in the limit $\tau \rightarrow 0$.

The new equations of motion read then as:
\beqa
 \partial_t I_j(\sigma)&=&  \{H_\tau,I_j(\sigma)\}= \partial_{\sigma}J^k \delta_{kj}\label{newone}\\ 
\partial_t J^j(\sigma)&=& \{H_\tau,J^j(\sigma)\}=\partial_{\sigma}I_k \delta^{kj}  - {\epsilon^{\ jl}}_{k}I_lJ^k. \label{newtwo}
\eeqa
which coincide with   Eqs. \eqn{eomI}, \eqn{eomJ}. 
Let us notice here that the same deformed algebra, namely the affine Lie algebra of $SL(2,\C)$ or $SO(4)$, according to $\tau$ being imaginary or real,  has been considered in \cite{klimcikdefo} with the main difference that in the latter case the author gets a true deformation of the dynamics, whereas in our case we have an alternative description of one and the same dynamics. As anticipated in the introduction, this should not be surprising, since the cotangent  space $T^*SU(2)$ and the phase space $SL(2;\C)$ are symplectomorphic. 

Let us rescale the fields according to 
\be
\frac{I}{(1-\tau^2)}\rightarrow I\;\;\;\; \frac{J}{(1-\tau^2)}\rightarrow J \label{redef}
\ee
so  that the Poisson algebra becomes 
\beqa
\{I_i(\sigma),I_j(\sigma')\}&= &{\epsilon_{ij}}^k I_k(\sigma)\delta(\sigma-\sigma'),   \label{modipoi1}\\
\{I_i(\sigma),J^j(\sigma')\}&= &J^k(\sigma){ \epsilon_{ki}}^j  \delta(\sigma-\sigma')-\delta_{i}^j\delta'(\sigma-\sigma'),  \label{modipoi2} \\
\{J^i(\sigma),J^j(\sigma')\}& = & \tau^2 {\epsilon^{ij}}_k I_k(\sigma)\delta(\sigma-\sigma') \label{modipoi3}
\eeqa
while  the rescaled Hamiltonian becomes identical to the undeformed one \eqn{undefoH}. 
Once identified the Lie algebra here described by the deformed Poisson brackets, one can define new generators which make it easier to recognize the bi-algebra structure on it. As in the finite dimensional case,  we keep the  generators of  $\mathfrak{su}(2)(\mathbb{R})$ unmodified  and consider the  linear combination:
 \be
 K^i(\sigma)=J^i(\sigma)-i\tau {\epsilon^{li3}}I_l(\sigma). \label{Kgen}
 \ee
From  the deformed Poisson brackets  \eqn{modipoi1}-\eqn{modipoi3} it is possible to derive  the Poisson brackets of the new generators:
%\begin{equation}
%\begin{split}
%& \{K_i(\sigma),K_j(\sigma')\}= \{J_i(\sigma)-\epsilon_{ik3}I_k(\sigma),J_j(\sigma')-\epsilon_{jl3}I_l(\sigma')\}= \\
%&=  \{J_i(\sigma),J_j(\sigma')\}-i\tau\epsilon_{jl3}\{J_i(\sigma),I_l(\sigma')\}-i\tau \epsilon_{ik3}\{I_k(\sigma),J_j(\sigma')\}+ \\
%& \quad +\tau^2 \epsilon_{ik3}\epsilon_{jl3}\{I_k(\sigma),I_l(\sigma')\}= \\
%& = (1-\tau^2)\tau^3\delta(\sigma-\sigma')[(-\epsilon_{ijk}+\epsilon_{is3}\epsilon_{jl3}\epsilon_{slk})\tau I_k(\sigma')-i(\epsilon_{il3}\epsilon_{ljk}-\epsilon_{jl3}\epsilon_{lik})J_k(\sigma')],  
%\end{split}
%\end{equation}
%so we can write 
\begin{equation}
%\begin{split}
\{K^i(\sigma),K^j(\sigma')\}
%&  (1-\tau^2)\tau^3\delta(\sigma-\sigma')[-\tau \epsilon_{ijl}\epsilon_{lk3}\epsilon_{ks3}I_s(\sigma')+i\epsilon_{ijl}\epsilon_{lk3}J_k(\sigma')]= \\
= i \tau\epsilon^{ijl}\epsilon_{l3k}K^k(\sigma') \,  \delta(\sigma-\sigma')\label{poisk}
%\end{split}
\end{equation}
showing that the $K$'s span the $\mathfrak{sb}(2,\mathbb{C})(\mathbb{R})$ Lie algebra, with structure constants ${f^{ij}}_k=\epsilon^{ijl}\epsilon_{l3k}$, while for the mixed Poisson brackets one finds:  
\beqa
 &&\{I_i(\sigma),K^j(\sigma')\} =\{I_i(\sigma),J^j(\sigma')-i\tau \epsilon^{jl3}I_l(\sigma')\} \nonumber\\
&&\;\;=\left(K^k(\sigma'){\epsilon_{ki}}^j-i \tau   I_k(\sigma') \epsilon^{kjs}{\epsilon_{s3i}}\right) \delta(\sigma-\sigma')
- \delta_{i}^j\delta'(\sigma-\sigma')
\label{poiski}
\eeqa
where  we recognize again the structure constants of the Lie algebra $\mathfrak{sb}(2,\C)$, $ \epsilon^{kjs}{\epsilon_{s3i}}={f^{kj}}_i$.  Notice that, in deriving the Poisson algebra  above one has to use the Jacobi identity for the structure constants of $SU(2)$ with one index equal to 3
$$
\epsilon_{qs3}\epsilon_{sji}+\epsilon_{is3}\epsilon_{qjs}+\epsilon_{js3}\epsilon_{iqs}=0
$$
yielding
\be
{f^{qi}}_j= \epsilon^{qis}\epsilon_{js3}= -{\epsilon^{qs}}_{3}{\epsilon_{sj}}^i-{\epsilon^{is}}_{3} {\epsilon^{q}}_{js}.
\ee
In this way,  the Lie algebra $\mathfrak{c}_2\equiv\mathfrak{sl}(2,\mathbb{C})(\mathbb{R})$ has been expressed as $\mathfrak{c}_2 =\mathfrak{su}(2)(\mathbb{R}) 
\Join\mathfrak{sb}(2,\mathbb{C})(\mathbb{R}),$ up to a central extension with central charge equal to $-1$, i.e. just like the affine algebra associated with the Drinfel'd double of the Lie algebra $\mathfrak{su}(2)$ considered at the beginning.\\
To summarize, upon rewriting the alternative Hamiltonian \eqn{modiha} in terms of the new generators, the $SU(2)$ chiral model is completely described  by the one-parameter family of Hamiltonian functions 
\begin{equation}
H_{\tau}=\frac{1}{2}\int_{\mathbb{R}} \mathrm{d}\sigma \  \left[I_sI_l\left( \delta_i^s\delta_j^l -\tau^2 {\epsilon^s}_{i3} {\epsilon^l}_{j3}\right)\delta^{ij}+ K^i K^j \delta_{ij} + 2i\tau {\epsilon}^{sl3} I_s K^q \delta_{l q}\right] \label{modiha2}
\end{equation}
with Poisson brackets given by:
\beqa
\{I_i(\sigma),I_j(\sigma')\}&= & {\epsilon_{ij}}^k I_k(\sigma)\delta(\sigma-\sigma') \label{IIcur} \\
\{K^i(\sigma),K^j(\sigma')\}&=&  i \tau{f^{ij}}_kK^k(\sigma') \delta(\sigma-\sigma')\label{KKcur}\\
\{I_i(\sigma),K^j(\sigma')\} &=&\left(K^k(\sigma'){\epsilon_{ki}}^j+i \tau   {f^{jk}}_i I_k(\sigma') \right) \delta(\sigma-\sigma')-\delta_{i}^j\delta'(\sigma-\sigma') \label{IKcur}
\eeqa
yielding the interesting  result that the Principal Chiral Model with compact target space may be described in terms of a non-compact current algebra. 
This result can be traced back to   the fact that the cotangent bundle of the group $SU(2)$ is symplectomorphic to the group $SL(2,\C).$ 
We shall see in the next section that this is not the case for the cotangent bundle of the  dual group of $SU(2)$.

Remarkably, the Hamiltonian \eqn{modiha2} may be rewritten in terms of a Riemannian metric which we choose to denote as an  inverse metric, ${{\mathcal H}_\tau}^{-1}$, for reasons that will be clear in a moment. By introducing:
\be \label{Htau-1}
{{\mathcal H}_\tau}^{-1}=
\begin{pmatrix}
 h^{ij}(\tau) &i\tau \epsilon^{ip3}\delta_{pj} \\
  i\tau \delta_{ip}\epsilon^{jp3} &  \delta_{ij}
\end{pmatrix}
\ee
where it has been defined, for  future convenience:
\be
h^{ij}(\tau)= \delta^{ij}-\tau^2 \epsilon^{ia3}\delta_{ab} \epsilon^{jb3} \label{htau}
\ee
 one has indeed:
\begin{equation}
H_{\tau}=\frac{1}{2}\int_{\mathbb{R}} \mathrm{d}\sigma \  \left[I_sI_l ({{\mathcal H}_\tau}^{-1})^{sl} + K^s K^l ({{\mathcal H}_\tau}^{-1})_{sl} + K^s I_l  {{( {{\mathcal H}_\tau}^{-1} )}_s}^{l} + I_s  K^l{ ({{\mathcal H}_\tau}^{-1})^{s}}_l \right].  \label{modiha3}
\end{equation}
Let us observe that the metric ${{\mathcal H}_\tau}^{-1}$  coincides with the inverse of $\mathcal{H}$ 
 defined in \eqn{nondegG} for $\tau=-i$, while $h^{ij}(\tau)\underset{\tau=\pm i}\rightarrow h^{ij}$ of Eq. \eqn{hij}. Moreover  one has:
\be
h_{ij}(\tau)= \delta_{ij}+\frac{\tau^2}{1-\tau^2} \epsilon_{ia3}\delta^{ab} \epsilon_{jb3} \label{invhtau}
\ee
with $h_{ij}(\tau)$  the inverse metric of $h^{ij}(\tau)$.
In terms of  the compact   notation ${I}_J= ( I_j, K^j)$, one can rewrite the Hamiltonian as:
\be\label{hcompactau}
H_{\tau}=\frac{1}{2}\int_{\mathbb{R}} \mathrm{d}\sigma \, { I}_L ({{\mathcal H}^{-1}_\tau})^{LM} { I}_M.
\ee

%To $H_{\tau}$  we can associate a Lagrangian by means of an inverse Legendre transform. 
%We have, for  the generalized velocities
%\be
%V^i(\sigma) = \frac{ \delta H}{\delta I_i (\sigma)}= ( h^{il}(\tau) I_l +  {{\mathcal H}^{-1}_\tau}^{i}_l K^l)
%= \delta^{is} I_s + i\tau \epsilon^{ia3}\delta_{al} J^l
%\ee
%so that the Lagrangian reads
%\be\label{Ltau}
%L_\tau&=& \frac{1}{2} \int d\sigma \left[-K^q {\delta_{qp}(\tau)} K^p +  V^q \delta_{qs}{h^{sr}({\tau})}\delta_{rp} V^p -2 i {\tau}  V^q {h_{qs}(\tau)} \epsilon^{sa3}\delta_{ap} K^p\right]
%\ee
%it being 
%$$ i\tau h_{qs}(\tau)  {\epsilon}^{sa3} \delta_{ap}= \frac{i\tau}{1-\tau^2}{\epsilon}_{qp3}$$
%with 
%\be
%h_{ij}(\tau)= \delta_{ij}+\frac{\tau^2}{1-\tau^2} \epsilon_{ia3}\delta^{ab} \epsilon_{jb3}. \label{invhtau}
%\ee
%the inverse metric of $h^{ij}(\tau)$. 
%with 
%\be \label{ktau}
 %{k_{qp}({\tau})}= \delta_{qp}+ \frac{2\tau^2}{1-\tau^2} \epsilon_{qs3}\delta^{sr}  \epsilon_{qs3}
 %\ee
 %being the inverse of the matrix $ {h_{\tau}}^{ij} = \delta^{ij}- \tau^2\epsilon^{is3} \delta_{sl} \epsilon^{j l3}$ appearing in \eqn{Htau-1}. %The Euler Lagrange equations are readily obtained to be
% \beqa
 %correctly reproducing the Lagrangian \eqn{unmodlag} in the limit $\tau\rightarrow 0$, being $K\rightarrow J, V\rightarrow A$.
 
Thus, summarizing the results of  this section, we have  a whole  family of models, labelled by the parameter $\tau$,   which are related (and indeed equivalent) to the standard $SU(2)$ chiral model  by the linear transformation \eqn{Kgen}, which     can be checked  to be a $O(3,3)$ transformation. %\textcolor{red}{questa e' un'ipotesi, andrebbe provato}. 
This transformation is a symmetry of the dynamics because it maps solutions into solutions. 
 
\subsubsection{Poisson-Lie structure}\label{poistruc}

 The PCM, in the formulation given by the Hamiltonian in Eq. \eqn{modiha3}, together with the Poisson algebra \eqn{IIcur}-\eqn{IKcur}, is a Poisson-Lie sigma-model according to the following analysis.

Keeping in mind that  $K^{i}, I_{i}$ are coordinate functions for the target phase space of the model,  $SU(2)\ltimes \mathfrak{\tilde{g}}$,  with $K^{i}$ base coordinates and $I_{i}$ fiber coordinates, we associate to $K^i$ the Hamiltonian vector fields \eqn{hamvecfi}
\be
X_{K^{i}} := \{  \cdot, K^{i} \} \label{hamv}
\ee
spanning the fibers which are isomorphic to   the vector space $\R^3$. Because of the non-trivial Poisson bracket \eqn{KKcur}, the latter becomes a non-Abelian algebra according to the following (cfr. Eqs. \eqn{havefialgebra1}-\eqn{havefialgebra3}):
\be
[X_{K^{i}}, X_{K^{j}} ] =  X_{\{ K^{i}, K^j \}} = i \tau f^{ij}_{k} X_{K^{k}}  \,\,\, .
\ee
Hence, we obtain  the dual Lie algebra $\mathfrak{sb}(2, \C)$, and  in the limit $\tau\rightarrow 0$   we recover the Abelian structure of the starting model over $T^*SU(2)$. 
A dual formulation of this property can be given in terms of the Hamiltonian vector fields associated with the currents $I_i$, say $X_i$. By repeating the analysis above, they can be seen to close the Lie algebra of $\mathfrak{su(2)}$, hence, they can be regarded  as one-forms over the dual Lie algebra, which has become non-Abelian, according to the computation above. We have then:
\be
dX_i ( \tX^{j}, \tX^{k}) = - X_{i} ( [\tX^{j}, \tX^{k}]) = - f_{i}^{\,\, jk} 
\ee
reproducing, in this way, the commonly used definition of Poisson-Lie structure.

\subsubsection{A family of Born geometries}\label{Born}
We have just seen in Sect. \ref{hamformul} how the deformation of the Poisson algebra $\mathfrak{c}_1=\mathfrak{su}(2)(\R) \ltimes \mathfrak{a}$ into $\mathfrak{c}_2 = \mathfrak{sl}(2, \mathbb{C})$ induces an alternative formulation of the Hamiltonian dynamics of the Principal Chiral Model with target space $SU(2).$ In this formulation we have seen the Riemannian metric $\mathcal{H}_\tau^{-1}$ \eqn{Htau-1} emerging in the definition of the alternative Hamiltonian $H_\tau$ \eqn{modiha3}. 

In order to understand the geometric meaning of such metric, let us take a step back to the original Hamiltonian $H$ \eqn{undefoH}. We can write the undeformed Hamiltonian as
\begin{equation}\label{compactH}
H=\frac{1}{2}\int_{\R} \mathrm{d} \sigma \ I_I \ (\mathcal{H}_0^{-1})^{IJ}\ I_J,  
\end{equation}
where $I_I=(I_i, J^i)$ are components of the current 1-form on $T^*SU(2)$ and 
\begin{equation} \label{hzero}
{(\mathcal{H}_0^{-1})}^{IJ}=
\begin{pmatrix}
\delta^{ij} & 0 \\
0 & \delta_{ij}
\end{pmatrix}
\end{equation} 
is a Riemannian metric on $T^*SU(2).$
In other words, the  Hamiltonian description of the Principal Chiral Model on $SU(2)$ naturally involves the Riemannian metric $\mathcal{H}_0^{-1}$ on the cotangent bundle. 
%From Eq. \eqn{compactH} it is evident that the Hamiltonian is a scalar under $O(3,3)$ transformations. 

Interestingly, the metric \eqn{hzero} can be interpreted as one of the structures defining a left-invariant Born geometry on $T^*SU(2)$. In the present case the transformation defining a Born geometry, as detailed below,  acts  as  an $O(3,3)$  transformation of  the target phase  $T^*SU(2)$.

The concept of Born reciprocity giving rise to Born geometries has been  first introduced, up to our knowledge, by Freidel and collaborators  in  \cite{Borngeomf}, in order to   provide a new point of view on string
theory in which spacetime is a derived dynamical concept. We shall see that the family of models which we have described in the previous section can be related with   such interpretation, with  the phase space of the chiral model regarded as dynamical.   Born reciprocity is thus implemented  as a choice of a Lagrangian submanifold of the phase space, in our case governed by the parameter $\tau$, and amounts to a generalization of T-duality. In this approach the phase space of the model can be understood in terms of dynamical bi-Lagrangian manifolds whose  geometric structure is an example of a  Born geometry. Let us notice that, in our case, such a bi-Lagrangian manifold happens to be a Drinfel'd double as well, with an interesting overlap between the two structures.

To this, let us  start by recalling  that $T^*SU(2)$ is a Drinfel'd double with Lie algebra $\mathfrak{su}(2) \ltimes \R^3.$  Such Lie algebra has a natural (symmetric, non-degenerate) pairing $\braket{\cdot \ ,\ \cdot }$ such that $\mathfrak{su}(2)$ and $\R^3$ are maximally isotropic subspaces with respect to it. Moreover, $\mathfrak{su}(2) \ltimes \R^3$ can be seen as a split vector space $\mathfrak{su}(2) \oplus \R^3,$ thus it can be naturally endowed with a para-complex structure $\kappa,$ i.e. $\kappa \in \mathrm{End}(\mathfrak{su}(2) \ltimes \R^3)$ such that $\kappa^2 =\mathds{1}$ with $\mathfrak{su}(2)$ eigenspace of $\kappa$ associated with the eigenvalue $+1$ and $\R^3$ eigenspace associated with the eigenvalue $-1.$ The structures  $\braket{\cdot \ ,\ \cdot }$ and $\kappa$ satisfy a compatibility condition $$\braket{\kappa(\xi), \psi}=-\braket{\kappa(\psi), \xi}, \hspace{1cm} \forall \xi, \psi \in \mathfrak{su}(2)\ltimes \R^3,$$ which defines a two-form $\omega$ on $\mathfrak{su}(2) \ltimes \R^3.$ Summarizing, $(\braket{\cdot \ ,\ \cdot }, \ \kappa)$ related by the above compatibility condition define a para-Hermitian structure on  $\mathfrak{su}(2) \ltimes \R^3.$

Since $T(T^*SU(2)) \cong T^*SU(2) \times  (\mathfrak{su}(2) \ltimes \R^3),$ we may read the structures $(\braket{\cdot \ , \ \cdot}, \kappa)$ as defined pointwise on $T^*SU(2),$ giving, respectively,  a left-invariant $O(3,3)$ metric $\eta$ and an endomorphism $\kappa$ of $T(T^*SU(2))$ such that $\kappa^2=\mathds{1}$ which has $TSU(2)$ (eigenvalue $+1$) and $T\R^3$  (eigenvalue $-1$) as eigenbundles 
\footnote{One may also say that $\pi : T^*SU(2)\rightarrow SU(2)$ is foliated by $SU(2)$ and $\R^3.$ Note that, in this case, the foliation $\R^3$ can be also seen as given by the canonical vertical subbundle $V=ker(d \pi)$ of $T(T^*SU(2))$ defined as the kernel of $\mathrm{d} \pi: T(T^*SU(2))\rightarrow TSU(2).$ The other foliation may be obtained by a choice of the horizontal distribution such that horizontal vectors are left-invariant with respect to $SU(2)$ , i.e. splitting the canonical short exact sequence $$0 \rightarrow V \rightarrow T(T^*SU(2))\rightarrow \pi^{\star}(TSU(2))\rightarrow 0$$  with the proper horizontal lift of left-invariant vector fields. A para-complex structure is naturally associated with such splitting.}
 (with a  slight abuse of notation for $\kappa$). Again, one has a so-called fundamental two-form $\omega=\eta \kappa$ on $T^*SU(2)$ coming from the compatibility of $\eta$ and $\kappa.$
  
In order to understand the relation between the left-invariant para-Hermitian structure $(\eta, \kappa)$ and the Riemannian metric  \eqn{hzero}, let us consider the (global) basis $\{\alpha^i, \varphi_i\}$ of left-invariant 1-forms on $T^*SU(2)$ with dual left-invariant vector fields $\{X_i, Y^i\}.$ Then, the para-Hermitian structure $(\eta, \kappa)$ on $T^*SU(2)$ can be written as 
\begin{align}
\eta &= \alpha^i \otimes \varphi_i + \varphi_i \otimes \theta^i \ , \\
\kappa &= X_i \otimes \alpha^i -Y^i \otimes \varphi_i \ .
\end{align}
The fundamental two-form then reads as:
\be
\omega= \eta \kappa= \varphi_i \wedge \alpha^i
\ee
and the Riemannian metric \eqn{hzero} is written as $$\mathcal{H}_0= \delta^{ij} \varphi_i \otimes \varphi_j + \delta_{ij} \alpha^i \otimes \alpha^j.$$ Note that the left-invariant Riemannian metric $\mathcal{H}_0$ is the unique left-invariant metric such that left-invariant vector fields are orthonormal. So its appearance in the Hamiltonian is completely natural in the context of Lie groups. Therefore, from the above expressions it is easy to verify that: $$\eta^{-1} \mathcal{H}_0=  \mathcal{H}_0^{-1} \eta \hspace{2cm} \omega^{-1} \mathcal{H}_0= - \mathcal{H}_0^{-1} \omega.$$
These are the defining relations for the Born structure $(\eta, \kappa, \mathcal{H}_0)$ on $T^*SU(2).$ This is the canonical Born geometry induced by the Drinfel'd double structure, see \cite{Borngeomf}, \cite{marottaszabo} for details.

The deformed Hamiltonian $H_\tau$ also gives a Riemannian metric on $T^*SU(2)$ and we shall see that such metric $\mathcal{H}_\tau$ is a $B$-transformation of the metric $\mathcal{H}_0.$  

Let us consider the $\tau$-dependent $B$-transformation 
\be \label{onepara}
e^{B(\tau)}= 
\begin{pmatrix}
\mathds{1} & i\tau B \\
0 & \mathds{1}
\end{pmatrix}
\in O(3,3) 
\ee
such that the components of the tensor $B$ are given by $B^{ij}= \epsilon^{ij3}$
% \footnote{One may regard it as $B= \epsilon^{ij3}X_j \otimes \tilde{\theta}_i.$} and $\tau \in \mathbb{C}.$ Note that, for $i\tau$ real, the $B$-transformations in the form \eqn{onepara} close a \textcolor{blue}{one-parameter} subgroup of $O(3,3),$ with the identity given by $e^B(0) = \mathds{1}$ and the inverse obtained replacing $i\tau$ with $-i\tau.$

The Riemannian metric $\mathcal{H}_\tau,$ inverse of \eqn{Htau-1}, is obtained by the  $B$-transformation acting on  $\mathcal{H}_0:$  
\be\label{Htrans}
\mathcal{H}_\tau= \bigl(e^{-B(\tau)}\bigr)^t \mathcal{H}_0 e^{B(\tau)},
\ee
i.e. it has components:
\be 
{({\mathcal H}_\tau})_{IJ}=
\begin{pmatrix}
 \delta_{ij}  & i\tau \delta_{ip}\epsilon^{jp3}  \\
i\tau \epsilon^{ip3}\delta_{pj} & \delta^{ij}- \tau^2\epsilon^{is3} \delta_{sl} \epsilon^{j l3}
\end{pmatrix}.
\ee

Furthermore, the left-invariant para-Hermitian structure $(\eta, \kappa)$ is transformed under $e^{B(\tau)}.$ In particular, the only structure which changes under such transformation is the para-complex structure $\kappa,$ i.e. 
\be
\kappa_{\tau}= e^{B(\tau)} \kappa e^{-B(\tau)}
\ee  
with  $\kappa_\tau$  still compatible with $\eta,$ so that the fundamental two-form becomes $\omega_{\tau}= \eta \kappa_\tau.$ 
In matrix form, the new almost para-Hermitian structure reads as:
\be
\kappa_{\tau}=
\begin{pmatrix}
\mathds{1} & 2 i\tau B \\
0 & \mathds{1}
\end{pmatrix}
\hspace{1cm} \eta=
\begin{pmatrix}
0 & \mathds{1} \\
\mathds{1} & 0
\end{pmatrix}
\hspace{1cm} \omega_\tau=
\begin{pmatrix}
0 & \mathds{1} \\
\mathds{1} & 2i \tau \eta B
\end{pmatrix}
\ee
where $\eta B \in \Gamma(\wedge^2 T^* \R^3).$ Note that the new almost para-Hermitian structure still has $TSU(2)$ as eigenbundle while $T\R^3$ is transformed in a non-involutive distribution $V_{\tau}$ whose sections are generated by vector fields in the form ${\bar Y}^i= Y^i + i \tau \epsilon^{ij3}X_j.$
We can easily check that the metric $\mathcal{H}_\tau$ gives a Born structure on $T^*SU(2)$ together with $(\eta, \kappa_\tau),$ for each value of the parameter $\tau.$ 
%In this sense we obtain a family of Born structures from the deformation of the current algebra.

We finally show that the the deformed current algebra defined in Eqs. \eqn{IIcur}-\eqn{IKcur}  is obtained via the same  $B$-transformation of the Poisson current algebra of the fields $J^i, I_i$ . The latter  can be stated in  terms of the Poisson bivector field:
\be
\Lambda = \int \dd\sigma \dd\sigma' ~ \Lambda^{IJ}(\sigma,\sigma') { X_I(\sigma)}\wedge { X_J(\sigma')}
\ee
with 
\beqa
\Lambda_{ij}(\sigma,\sigma')&=& {\epsilon_{ij}}^k I_k \delta(\sigma-\sigma')\\
\Lambda^{ij}(\sigma,\sigma')&=& 0\\
{\Lambda_i}^j (\sigma,\sigma')&=& J^k(\sigma) {\epsilon_{ki}}^j \delta(\sigma-\sigma') -{\delta_i}^j \del_\sigma \delta(\sigma-\sigma')\\
{\Lambda^i}_j (\sigma,\sigma')&=& - J^k(\sigma) {\epsilon_{kj}}^i  \delta(\sigma-\sigma') +{\delta^i}_j \del_\sigma \delta(\sigma-\sigma') \,\,\, .
\eeqa
Thus, the $B$-transformed Poisson structure reads as:
\be
\Lambda'_{\tau}= e^{B(\tau)}\Lambda (e^{-B(\tau)})^t
\ee
namely
\be
{\Lambda'_{\tau}}^{IJ}=
\begin{pmatrix}
 { \delta^i}_j &i\tau B^{ij}\\
 0& { \delta_i}^j
\end{pmatrix}
\begin{pmatrix}
  0 &{\Lambda^j}_k \\
{\Lambda_j}^k & \Lambda_{jk}
\end{pmatrix}
\begin{pmatrix}
 { \delta_k}^l &0\\
-i\tau B^{kl} &  {\delta^k}_l
\end{pmatrix}
\ee
so that we may read the $B$-transformation $e^{B(\tau)}$ as a Poisson map between $\Lambda$ and $\Lambda'_{\tau},$ both bivector fields on $T^*SU(2).$ Note that $\Lambda$ is the Poisson structure on $T^*SU(2)$ obtained from the canonical (left-invariant) symplectic structure, as shown in Appendix \ref{appA}.
If we simultaneously  rotate the fields according to 
\be\label{fieldtrans}
\begin{pmatrix}
 J'^i\\
I'_i\end{pmatrix}= 
\begin{pmatrix}
 { \delta^i}_j & i\tau B^{ij} \\
 0 & { \delta_i}^j
\end{pmatrix}
\begin{pmatrix}
 J^j\\
I_j\end{pmatrix}
\ee
which is nothing but the $O(3,3)$ transformation \eqref{Kgen},  we reproduce the current algebra of the fields $I, \ K,$ i.e. \eqn{IIcur}, \eqn{KKcur} and \eqn{IKcur}, upon identifying $J'$ with $K$.

%Let us also notice that, for $\tau= -i$ this exactly the Lie algebra linear transformation that we have performed on the boost generators of the algebra $\mathfrak{sl}(2,\C)$ in Eq. \eqn{etilde} in order to map $\mathfrak{sl}(2,\C)$ to the Drinfel'd double $\mathfrak{su}(2)\bowtie \mathfrak{sb}(2,\C)$. 
Finally,  performing the transformations \eqn{Htrans}, \eqn{fieldtrans} on the Hamiltonian \eqn{compactH} we recover the expression  
\eqn{hcompactau}.

Therefore we can conclude by saying that the family of equivalent Hamiltonian descriptions of the $SU(2)$ PCM, first found in \cite{R89,R892}, can be understood in terms of a one-parameter family of Born geometries for the target phase space $T^*S^3$,  corresponding, for each choice of the parameter $\tau$,  to a specific splitting of phase space, with the value $\tau=0$  the canonical splitting.
%, whereas $i\tau=\pm1$ selects as Lagrangian submanifolds the two dual subgroups $SU(2)$ and $SB(2,\C)$ of the Drinfel'd double $SL(2,\C)$. 

\section{ Poisson-Lie dual models}\label{PLD}
From the Hamiltonian formulation of the $SU(2)$ chiral model we have seen that it is possible to describe the dynamics in terms of the centrally extended current algebra $\mathfrak{c}_2=\mathfrak{sl}(2,\C)(\R)$.  Therefore,   as we have done for the rigid rotor,  we shall look for a model whose target space is the dual group of $SU(2)$. As previously anticipated, we shall see that  the duality relation between the two models defined on the manifold of Poisson-Lie dual groups, is much more natural in the context of field theory.  To this, it is worth recalling that the model described above is a Poisson-Lie sigma model, as we have shown in sec. \ref{poistruc}.

Let us consider the Poisson algebra $\mathfrak{c}_2$ (with central extension), represented by Eqs. \eqn{IIcur}-\eqn{IKcur} and let us introduce another imaginary parameter $\alpha$ in such a way to make the role of the  subalgebras $\mathfrak{su}(2)(\R)$ and $\mathfrak{sb}(2,\C)(\R)$ symmetric. 
We consider namely the two-parameters generalization of the algebra \eqn{IIcur}-\eqn{IKcur} 
\beqa
\{I_i(\sigma),I_j(\sigma')\}&= & i\alpha\;  {\epsilon_{ij}}^k I_k(\sigma)\delta(\sigma-\sigma') \label{IIcural} \\
\{K^i(\sigma),K^j(\sigma')\}&=&  i \tau{f^{ij}}_kK^k(\sigma') \delta(\sigma-\sigma')\label{KKcural}\\
\{I_i(\sigma),K^j(\sigma')\} &=&\left(i\alpha  K^k(\sigma'){\epsilon_{ki}}^j+i \tau   {f^{jk}}_i I_k(\sigma') \right) \delta(\sigma-\sigma')-\delta_{i}^j\delta'(\sigma-\sigma') \label{IKcural}
\eeqa
which, in  the limit $i \tau\rightarrow 0$, reproduces the semi-direct sum $\mathfrak{su}(2)(\mathbb{R})\ltimes \mathfrak{a}$, while the limit $i\alpha\rightarrow 0$ yields $ \mathfrak{sb}(2,\C)(\mathbb{R})\ltimes \mathfrak{a}$. 
For all non zero  values of the two parameters,  the algebra is isomorphic to $\mathfrak{c}_2$, and, upon suitably rescaling the fields, one gets a two-parameter family of models, all equivalent to the Principal Chiral Model. 

Since the result might appear surprising at a first sight, let us show in detail how it works, by slightly generalizing the procedure of subsection \ref{hamformul} .  The goal is to show  that the  dynamics that is derived from the algebra \eqn{IIcural}-\eqn{IKcural}, together with a suitable Hamiltonian, is equivalent to the dynamics that follows from Eqs.  \eqn{IIbr}-\eqn{JJbr} with the Hamiltonian \eqn{undefoH}. As an intermediate step, one has to rescale the fields $I$ and $K$ as follows:
\be
\bar I_j= \frac{I_j}{i\alpha}\;\;\; \bar K^j= i\alpha K^j
\ee
which yield
\beqa
\{\bar I_i(\sigma),\bar I_j(\sigma')\}&= & \;  {\epsilon_{ij}}^k \bar I_k(\sigma)\delta(\sigma-\sigma') \label{IIcurral} \\
\{\bar K^i(\sigma),\bar K^j(\sigma')\}&=&  (i \tau\, i\alpha) {f^{ij}}_k \bar K^k(\sigma') \delta(\sigma-\sigma')\label{KKcurral}\\
\{\bar I_i(\sigma),\bar K^j(\sigma')\} &=&\left(  \bar K^k(\sigma'){\epsilon_{ki}}^j+(i \tau\, i\alpha)  {f^{jk}}_i \bar I_k(\sigma') \right) \delta(\sigma-\sigma')-\delta_{i}^j\delta'(\sigma-\sigma'). \label{IKcurral}
\eeqa
The latter is identical to  the algebra \eqn{IIcur}-\eqn{IKcur} , upon introducing $\bar \tau$, s.t. $i\bar\tau= i\tau \,i\alpha$. Then we rescale and rotate the fields, analogously  to what has been previously done, according to:
\be 
\hat  I_i= (1-  \alpha^2\tau^2) \frac{I_i}{i\alpha}\;\;\;\; \hat J^i= (1-  \alpha^2\tau^2)( i\alpha K^i + i \tau \epsilon^{li3} I_l)
\ee
so that the latter obey the Poisson algebra
\begin{align} 
\{\hat I_i(\sigma),\hat I_j(\sigma')\}= & (1-\alpha^2\tau^2){\epsilon_{ij}}^k \bar I_k(\sigma)\delta(\sigma-\sigma'),   \label{modipois1}\\
\{\hat I_i(\sigma),\hat J^j(\sigma')\}= & (1-\alpha^2 \tau^2)J^k(\sigma){ \epsilon_{ki}}^j  \delta(\sigma-\sigma')-(1-\alpha^2 \tau^2)^2\delta_{i}^j\delta'(\sigma-\sigma'),  \label{modipois2} \\
\{\hat J^i(\sigma),\hat J^j(\sigma')\}= & (1-\alpha^2 \tau^2)\tau^2 {\epsilon^{ij}}_k \hat I_k(\sigma)\delta(\sigma-\sigma') \label{modipois3}
\end{align}
that, together with the  modified Hamiltonian
\begin{equation}
H_{\tau,\alpha}=\frac{1}{2(1-\alpha^2 \tau^2)^2}\int_{\mathbb{R}} \mathrm{d}\sigma \  (\hat I_i  \hat I_j \delta^{ij}+\hat J^i \hat J^j \delta_{ij}) \label{modihanew}
\end{equation}
can be checked to yield the equations of motion
\beqa
 \partial_t \hat I_j(\sigma)&=&  \{H_{\tau, \alpha}, \hat I_j(\sigma)\}= \partial_{\sigma}\hat J^k \delta_{kj}\label{vnewone}\\ 
\partial_t \hat J^j(\sigma)&=& \{H_{\tau, \alpha}J^j(\sigma)\}=\partial_{\sigma}I_k \delta^{kj}  - {\epsilon^{\ jl}}_{k}I_lJ^k. \label{vnewtwo}
\eeqa
namely, the undeformed dynamics of the PCM as in Eqs. \eqn{eomI}, \eqn{eomJ}. In the limit $i \bar \tau \rightarrow 0$, the algebra and the Hamiltonian reduce to the original ones.  Notice that the factor $1-\alpha^2 \tau^2=1-\bar\tau^2$ is never zero for imaginary $\bar \tau$. 
\\
Once we have shown how to recast the  two-parameter algebra \eqn{IIcural}-\eqn{IKcural} in the form \eqn{modipois1}-\eqn{modipois3}, it is useful to express  the Hamiltonian \eqn{modihanew} in terms of the fields $I,K$. We get:
\begin{equation}
H_{\tau,\alpha}=\frac{1}{2}\int_{\mathbb{R}} \mathrm{d}\sigma \  \left[I_s I_l ( { { \mathcal H}_{\tau,\alpha}}^{-1} )^{sl} + K^s K^l ({{\mathcal H}_{\tau,\alpha}}^{-1})_{sl} + K^s I_l  {{( {{\mathcal H}_{\tau,\alpha}}^{-1} )}_s}^{l} + I_s  K^l{ ({{\mathcal H}_{\tau,\alpha}}^{-1})^{s}}_l \right].  \label{modiha4}
\end{equation}
with 
\be \label{Htaual-1}
{{\mathcal H}_{\tau,\alpha}}^{-1}=
\begin{pmatrix}
 \frac{h^{ij}(\bar\tau)}{(i\alpha)^2} &i\bar \tau \epsilon^{ip3}\delta_{pj} \\
  i\bar \tau \delta_{ip}\epsilon^{jp3} &  (i\alpha)^2\delta_{ij}
\end{pmatrix}
\ee
and $i\bar\tau$ previously defined. 
In terms of  the compact   notation ${I}_J= ( I_j, K^j)$, one can rewrite the Hamiltonian as:
\be\label{hcompactaual}
H_{\tau\alpha}=\frac{1}{2}\int_{\mathbb{R}} \mathrm{d}\sigma \, { I}_L ({{\mathcal H}^{-1}_{\tau,\alpha}})^{LM} { I}_M.
\ee
\\
Since the role of $I$ and $K$ is now symmetric, we can perform an  $O(3,3)$  transformation which exchanges the momenta ${I_i}$ with the  fields  ${K^i}$, thus obtaining a new two-parameter family of models, which legitimately deserve to be called  {\it duals} to the PCM. 
\\
 The $O(3,3)$ transformation 
\be
\tilde K(\sigma)={I(\sigma)}, \;\;\; \tilde I(\sigma)= K(\sigma)  \label{tildeuntilde}
\ee 
yields, when applied to the   Hamiltonian \eqn{modiha3}, the new Hamiltonian
\be\label{tilalpha}
\tilde H_{\tau,\alpha}= \frac{1}{2}  \int_\R\dd\sigma \  \left[\tilde K _s ( { { \mathcal H}_{\tau,\alpha}}^{-1} )^{sl} \tilde K_l  + \tilde I^s( { { \mathcal H}_{\tau,\alpha}}^{-1} )_{sl} \tilde I^l + 2 i{{( {{\mathcal H}_{\tau,\alpha}}^{-1} )}^s}_{l}  \tilde K_s \tilde I ^l \right] 
\ee
with Poisson algebra:
\beqa
\{\tK_i(\sigma),\tK_j(\sigma')\}&= & i\alpha\;  {\epsilon_{ij}}^k \tK_k(\sigma)\delta(\sigma-\sigma') \label{tKtKcur}\\
\{\tI^i(\sigma),\tI^j(\sigma')\}&=&  i \tau{f^{ij}}_k \tI^k(\sigma') \delta(\sigma-\sigma')\label{tItIcur}\\
\{\tK_i(\sigma),\tI^j(\sigma')\} &=&\left(i\alpha  \tI^k(\sigma'){\epsilon_{ki}}^j+i \tau   {f^{jk}}_i \tK_k(\sigma') \right) \delta(\sigma-\sigma')-\delta_{i}^j\delta'(\sigma-\sigma') \,\,\, . \label{tItKcur}
\eeqa
The Hamiltonian 
can be recast into the form 
\be
\tilde H_{\tau}=\frac{1}{2}  \int_\R\dd\sigma  \ \tI_I \ ({ \mathcal{H}}_{\tau,\alpha})^{IJ}\ \tI_J, \label{hamww}
\ee
%having posed
%\be \label{gensb}
%{\tilde{\mathcal H}}_{\tau}=
%\begin{pmatrix}
 %{{ h}^{\,ij} ({\tau}}) & i \tau \epsilon^{il3}\delta_{lj}  \\
%i\tau \delta_{il}\epsilon^{jl3} &\delta_{ij}
%\end{pmatrix}
%\ee
with  $\tI_J=(\tI^j, \tK_j )$. 
%%%%%%%%the following should be checked%%%%%%%%%%%%%%
%Alternatively, we can write:
%\beqa
%\label{moditha3}
%\tilde H_{\tau}&=&  \frac{1}{2} \int_\R\dd\sigma \  \left[\tilde J _s h^{sl}(\tau) \tilde J_l  + \tilde I^i(\delta_{ij} + \frac{\tau^2}{1-\tau^2} \epsilon_{ia3}\delta^{ab}\epsilon_{jb3}) \tilde I^j \right]\nn\\
%&=&\frac{1}{2} \int_\R\dd\sigma \  \left[\tilde J _s h^{sl}(\tau) \tilde J_l  + \tilde I^s h_{sl}(\tau) \tilde I^l \right]
%\eeqa
%where we have introduced the linear combination 
%\be
%\tilde J_s=\tilde K_s+ h_{sm}(\tau)  i\tau {\epsilon}^{ml3} \delta_{l q}\tilde I^q= \tK_s+\frac{i\tau}{1-\tau^2}{\epsilon}_{sq3}\tI^q \label{tilJ}
%\ee
%and the inverse metric $h_{ij}(\tau)$ given by  \eqn{invhtau} naturally appears.
%%%%%%%%%%%%%%%%%%%%%%%%%%%%%%%%%%%%%%%%%%%%%%%%
\\
From the Poisson algebra  \eqn{tKtKcur}-\eqn{tItKcur} we observe  that the new family of models, which we call DPCM (Dual Principal Chiral Models),  has target configuration space the group manifold of $SB(2,\C)$, spanned by the fields $\tK_i$, and momenta $\tI^i$ which span the fibers of the target phase space.  
%which, for the ridefinition \eqn{tilJ} becomes
%\beqa
%\{\tJ_i(\sigma),\tJ_j(\sigma')\}&= & \tJ_k\left( i\alpha\;  {\epsilon_{ij}}^k-\frac{2\tau^2}{1-\tau^2} {f^{kn}}_j \epsilon_{jn3}\right) \delta(\sigma-\sigma') \label{tJtJcur }\\
%\{\tI^i(\sigma),\tI^j(\sigma')\}&=&  i \tau{f^{ij}}_k \tI^k(\sigma') \delta(\sigma-\sigma')\label{tItIcur}\\
%\{\tJ_i(\sigma),\tI^j(\sigma')\} &=&\left(i\alpha  \tI^k(\sigma'){\epsilon_{ki}}^j+i \tau   {f_i}^{jk}\tK_k(\sigma') \right) \delta(\sigma-\sigma')-\delta_{i}^j\delta'(\sigma-\sigma') \label{tItKcur}
%\eeqa
\\
In strict analogy with what we have found previously, we could repeat step by step the analysis performed in section \ref{poistruc} and conclude that the DPCM are Poisson-Lie sigma models according to the definition we have given. Moreover, the two families are dual to each other by construction.
\\
To conclude this section, let us observe that, in the limit  $\alpha\rightarrow 0$ the dual current algebra collapses to the semidirect sum $\mathfrak{sb}(2,\C)(\mathbb{R})\ltimes \mathfrak{a}$,  but  the Hamiltonian \eqn{hamww} becomes singular. 

In the next section we will approach  the problem from a Lagrangian perspective, starting directly with a natural action defined on the Poisson-Lie dual of $SU(2)$.

\subsection{The Lagrangian approach}\label{duallagsec}
Following the approach that we have already used for the rigid rotor, it is  natural, within the Lagrangian approach,  to introduce  fields $\tilde g:(t,\sigma)\rightarrow SB(2,\C)$ and one-forms valued in the Lie algebra $\mathfrak{sb}(2,\C)$, in terms of which a natural Lagrangian can be defined on the Lie-Poisson dual to $SU(2)$.  The Hamiltonian will then be  obtained by Legendre transform, together with a Poisson algebra which, not surprisingly, will result to be  isomorphic to $\mathfrak{c}_3=\mathfrak{sb}(2,\C)(\mathbb{R})\ltimes \mathfrak{a}$. This new Hamiltonian will be related to  the two-parameter family of dual models introduced above, through a $B$-transformation.
 
Let us  look at the Lagrangian approach in some detail.

The action of the proposed model is a straightforward extension   of the one in Eq. \eqn{duac} to fields $\tilde\phi : (t,\sigma) \in  \R^{1,1}\rightarrow \tg\in   SB(2,\C)$, with  Lie algebra valued  left-invariant one-forms $\tg^{-1}\mathrm{d}\tilde g$ whose pull-back to $\R^{1,1}$ is given by:
\be
\tilde\phi^*(\tg^{-1}\mathrm{d}\tilde g)= (\tilde g^{-1}\partial_{t}\tilde g )_i \tilde e^i \,\mathrm{d}t+(\tilde g^{-1}\partial_{\sigma}\tilde g)_i  \tilde e^i\,\mathrm{d}\sigma.
\ee
We have then:
\be\label{dualchiact}
\tilde S=\frac{1}{2}\int_{\mathbb{R}^{1,1}} {\mathcal Tr} \left[ \phi^*(\tilde g^{-1}\mathrm{d}\tilde g) \wedge \h \phi^*( \tilde g^{-1}\mathrm{d}\tilde g)\right],
\ee
where, as in the finite-dimensional case, ${\mathcal Tr} $ stands for the non-degenerate product in the Lie algebra $\mathfrak{sb}(2,\C)$,  given by \eqn{nondegG},  and the Hodge star operator acts as $\h\dd t=\dd\sigma, \h\dd\sigma=\dd t$, yielding
\be
\tilde S= \frac{1}{2}\int_{\mathbb{R}^2}\mathrm{d}t\mathrm{d}\sigma\ \bigl[  (\tilde g^{-1}\partial_t\tg)_i  (\tilde g^{-1}\partial_t\tg)_j - (\tg^{-1}\partial_{\sigma}g)_i  (\tg^{-1}\partial_{\sigma}g)_j   \bigr]h^{ij} . \label{actdu}
\ee
%it being
%\be
%g^{-1}\partial_t g=(g^{-1}\partial_t g)^ie_i, \quad g^{-1}\partial_{\sigma}g=(g^{-1}\partial_{\sigma}g)^ie_i,
%\ee
As for the finite-dimensional case, the action functional is invariant under left $SB(2,\C)$ action.
 The Euler-Lagrange equations
\begin{equation}
h^{ij}\left(\partial_t(\tg^{-1}\partial_t \tg)_j-\partial_{\sigma}(\tg^{-1}\partial_{\sigma}\tg)_j\right)= {\sf L}_{\tX^i} \tilde L
% ={f^i}_{jk} \left(..\right).  
\end{equation}
%\textcolor{blue}{[COMMENT TO BE REMOVED: The last term replaces in this case $\frac{\del \tilde L}{\del \phi^i}$, $\phi^i$ being here the fields $g(\sigma)$]}
with $\tX^i(\sigma) $ the left-invariant vector fields over the group manifold and $\tilde L$ the Lagrangian, 
may be rewritten in terms of  an equivalent system of two first order partial differential equations, introducing, as for the $SU(2)$ principal model,  the currents\footnote{No factor  two is needed here because ${\mathcal Tr}(\te^i\te^j)=\delta^{ij}$}:
\begin{equation}
\tilde A_i=(\tg^{-1}\partial_t \tg)_i , \quad {\tJ}_i=(\tg^{-1}\partial_{\sigma}\tg)_i .
\label{currdu}
\end{equation}
The  Lagrangian becomes then:
\be
\tilde L= \frac{1}{2}\int_\R \dd\sigma (\tilde A_ih^{ij} \tilde A_j - \tJ_ih^{ij} \tilde J_k) \label{unmodualag}
\ee
and the equations of motion read:
\begin{align} 
h^{ij}(\partial_t \tilde A_j- \partial_{\sigma}\tJ_j)  = &{ f^{si}}_{l} h^{lj} (\tA_s \tA_j -\tJ_s\tJ_j),  
\label{equivdu1} \\
\partial_t \tJ = & \partial_{\sigma} \tilde A  -[\tilde A,\tJ],
 \label{equivdu2}
\end{align}
being the latter a condition for the  existence of  $\tg \in SB(2,\C)$ that admits the expression of the currents in the form \eqref{currdu}. 
At fixed $t$, all elements $\tg$ satisfying the boundary condition $\underset{\sigma\to\pm\infty}{ \lim} g(\sigma)=1$ form the  infinite-dimensional Lie group $SB(2,\C)({\R})\equiv \mathrm{Map}(\R,SB(2,\C))$, given by smooth maps $\tg: \sigma\in \mathbb{R}\rightarrow \tg(\sigma)\in SB(2,\C)$ which are constant at infinity. 
 
 At fixed time, the currents $\tJ$ and $\tilde A$ take values in the Lie algebra $\mathfrak{sb}(2,\C)(\mathbb{R})$ of functions from $\mathbb{R}$ to $\mathfrak{sb}(2,\C)$ that are sufficiently fast decreasing at infinity to be square-integrable.  Therefore the tangent bundle description of the dual dynamics can be given in terms of $(\tJ,\tilde A)$, with $\tilde A$ the left generalized velocities, while $\tJ$ playing the role of left configuration space coordinates.

%We choose these functions to be from $\mathbb{R}$ in order to have equal-time (Poisson) Lie bracket on  $\mathfrak{g}(\mathbb{R})$, as usual in field theory. Substantially, we can write the elements in this Lie algebra as $I= I^i (\sigma)e_i$ and $J=J^i(\sigma)e_i$.\\
%Let us state this in a proper form \cite{sinistro}.
%\theoremstyle{definition}
%\begin{definition}
%Let $G$ be a compact Lie group. The group $G(M)$, or $Map(M,G)$, associated with $G$ over a smooth manifold $M$ is the Lie group given by the smooth mappings $$g: M \ni x \rightarrow g(x) \in G,$$
%where the pointwise group law is the composition of functions $$g''(x)=g'(x)\circ g(x), \ \mathrm{with}\  g''(x),g'(x),g(x) \in G(M),$$ the inverse element is $g^{-1}=[g(x)]^{-1}$ and the unit element is the identity in $G$, $e(x)=e$. \\
%The group $G(M)$ is infinite-dimensional.
%\end{definition}
%%%%%%%%%%%%

\subsubsection{The Hamiltonian description}
Upon introducing left momenta
\be
\tI^i=\frac{\delta \tilde L}{\delta \ (\tg^{-1}\partial_t \tg)_i}=(\tg^{-1}\partial_t\tg)_jh^{ij}={\tilde A}_jh^{ij}
\end{equation}
and inverting for the generalized velocities, one obtains the Hamiltonian:
\be 
\tilde H=\frac{1}{2}\int_{\mathbb{R}} \mathrm{d}\sigma \tI^i \tI^j  h_{ij}+\tJ_i \tJ_j  h^{ij}= \frac{1}{2}\int_{\mathbb{R}} \mathrm{d}\sigma \tI_I ({{\tilde{\mathcal{K}}}_0}^{-1})^{IJ} \tI_J   \label{sbham}
\ee
with 
\be \label{metsb0}
{\tilde{\mathcal K}}_0=
\begin{pmatrix}
 { h}^{\,ij}  & 0  \\
 0 &  h_{ij}
\end{pmatrix}
\ee
and $\tI_J=(\tI^j,\tJ_j)$,  while the equal-time 
Poisson brackets can be derived in the usual way  from the action functional (see appendix \ref{appA})  to be 
\begin{align} 
\{\tilde{I}^i(\sigma),\tilde{I}^j(\sigma')\}= & {f^{ij}}_k \tilde{I}^k(\sigma)\delta(\sigma-\sigma'),  \label{poidu1} \\
\{\tilde{I}^i(\sigma),\tilde{J}_j(\sigma')\}= &\tilde{J}_k(\sigma) {f^{ki}}_j \delta(\sigma-\sigma') -\delta^i_{j}\delta'(\sigma-\sigma'),  \label{poidu2} \\
\{\tilde{J}_i(\sigma),{\tJ}_j(\sigma')\}= & 0  \label{poidu3}
\end{align}
yielding the equations of motion
\beqa
\del_t \tilde{I}^i(\sigma)&=&  \tI^s\tI^r {f^{ji}}_s h_{rj}  - \tJ_r \tJ_s {f^{si}}_j h^{rj}  + \delta^i_j\, h^{rj}\, \del_\sigma \tJ_r  \\
\del_t \tJ_i (\sigma)&=& \left(\tI^s \tJ_k {f^{kj}}_i + \delta^j_i \del_\sigma \tI^s\right)h_{sj} \,\,\, . 
\eeqa
The Poisson brackets \eqn{poidu1}-\eqn{poidu3} realize the current algebra 
 $\mathfrak{c_3}=\mathfrak{sb}(2,\mathbb{C})(\mathbb{R})\ltimes\mathfrak{a}$, which we have already regarded as the limit $i\alpha\rightarrow 0$ of the algebra \eqn{IIcur}-\eqn{IKcur}.
 %\footnote{Let us notice that, upon defining $\tilde W_i= \tJ_i - \frac{1}{2}{\epsilon}_{iq3}\tI^q$ 
  %we have an equivalent description in terms of the Hamiltonian 
  %\be 
  %\tilde H=\frac{1}{2} \int_\R\dd\sigma \  \left[\tilde W _i h^{ij} \tilde W_j  + \tilde I^ih_{ij} \tilde I^j \right]
  %\ee and Poisson algebra 
 %\beqa
%\{{\tilde W}_i(\sigma),\tilde W_j(\sigma')\}&= &{f^{kn}}_j \epsilon_{in3}\tilde W_k(\sigma)\delta(\sigma-\sigma') \label{defo1}\\
%\{\tI^i(\sigma),\tI^j(\sigma')\}&=& {f^{ij}}_k \tI^k(\sigma') \delta(\sigma-\sigma')\label{defo2} \\
%\{{\tilde W}_i(\sigma),\tI^j(\sigma')\} &=& \left( -\tI^n(\sigma')\epsilon_{kn 3} {f^{jk}}_i +  {f^{jk}}_i \tilde W_k(\sigma') \right) \delta(\sigma-\sigma')-\delta_{i}^j\delta'(\sigma-\sigma') \label{defo3}
%\eeqa
 %\label{note}
%}
 %, upon identifying $\tilde J$ with $\tilde K$.  

Similarly to the $SU(2)$ PCM,   the currents $(\tJ, \tI)$ may be identified with cotangent space left coordinates for $T^*SB(2,\C)(\R)$. 
However, differently from  $T^*SU(2)$, $T^*SB(2,\C)$ is not symplectomorphic to $SL(2,\C)$, the two spaces being topologically different to start with.   Therefore, certainly the model cannot be given an equivalent description in terms of an $SL(2,\C)(\R)$ algebra. 
Indeed,  it will be shown, in the next section, that  the  $SB(2,\C)$ PCM Hamiltonian obtained here through Legendre transform can be related to the DPCM models previously found,   through a $B$-transformation, but not its Poisson algebra.   %We claim therefore that the two models are dual to each other.  

\subsubsection{Dual Born geometry}
   
Following the same approach as in Section \ref{Born}, let us recall that a left-invariant para-Hermitian structure $(\tilde{\eta}, \tilde{\kappa})$ can be defined on $T^*SB(2,\C),$ as discussed for $T^*SU(2),$ starting from its Lie algebra $\mathfrak{sb}(2, \C) \ltimes \R^3.$ Thus $\tilde{\kappa}$ comes from the splitting of $\mathfrak{sb}(2, \C) \ltimes \R^3$ as a vector space and $\tilde{\eta}$ is obtained from the duality pairing.  The fundamental two-form of such structure is denoted by $\tilde{\omega}.$

According to what has been done in Section \ref{Born}, let us start from the metric ${\tilde{\mathcal{K}}}_0$. It is easily verified that 
it  is Riemannian, with determinant equal to 1 and such that
\be
{\tilde{\mathcal K}}^T_0 \tilde{\eta} {\tilde{\mathcal K}}_0=\tilde{\eta} \,\, .
\ee
We consider the $\beta$-dependent $B$-transformation 
\be \label{oneparaB}
e^{B(\beta)}= 
\begin{pmatrix}
\mathds{1} & i\beta B \\
0 & \mathds{1}
\end{pmatrix}
\in O(3,3) 
\ee
with   $B^{ij}= \epsilon^{ij3}$  as before and $\beta$ an imaginary parameter. 

The Riemannian metric $\tilde{\mathcal{K}}_\beta$, can be  obtained by the  $B$-transformation acting on  $\tilde{\mathcal{K}}_0:$  
\be\label{Ktrans}
\tilde{\mathcal{K}}_\beta= \bigl(e^{-B(\beta)}\bigr)^t \tilde{\mathcal{K}}_0 e^{B(\beta)},
\ee
yielding
\be \label{Kbeta}
{(\tilde{\mathcal K}_\beta})_{IJ}=
\begin{pmatrix}
 h^{ij}  & 2i\beta \epsilon^{il3} \delta_{lj}  \\
2i\beta  \delta_{il}\epsilon^{jl3}& \delta_{ij}- \epsilon_{il3}\delta^{lk}\epsilon_{jk3}(2\beta^2+\frac{1}{2}) 
\end{pmatrix} \,\,\, .
\ee

Furthermore, the left-invariant para-Hermitian structure $(\tilde{\eta}, \tilde{\kappa})$ is transformed under $e^{B(\beta)}$ with  
\be
\tilde{\kappa}_{\beta}= e^{B(\beta)} \tilde{\kappa} e^{-B(\beta)}
\ee  
 still compatible with $\tilde{\eta}$, so that the fundamental two-form becomes $\tilde{\omega}_{\beta}= \tilde{\eta} \tilde{\kappa}_\beta.$ 
In matrix form, the new almost para-Hermitian structure reads as:
\be
\kappa_{\beta}=
\begin{pmatrix}
\mathds{1} & 2 i\beta B \\
0 & \mathds{1}
\end{pmatrix}
\hspace{1cm} \eta=
\begin{pmatrix}
0 & \mathds{1} \\
\mathds{1} & 0
\end{pmatrix}
\hspace{1cm} \omega_\beta=
\begin{pmatrix}
0 & \mathds{1} \\
\mathds{1} & 2i \beta \eta B
\end{pmatrix}
\ee
where $\eta B \in \Gamma(\wedge^2 T^* \R^3).$ The new almost para-Hermitian structure still has $TSB(2,\C)$ as eigenbundle while $T\R^3$ is transformed in a non-involutive distribution $V_{\beta}$ whose sections are generated by vector fields in the form ${\bar Y}^i= {Y}^i + i \beta \epsilon^{ij3}X_j.$
%We can easily check that the metric $\mathcal{H}_\tau$ gives a Born structure on $T^*SU(2)$ together with $(\eta, \kappa_\tau),$ for each value of the parameter $\tau.$ 
%In this sense we obtain a family of Born structures from the deformation of the current algebra.

Let us compare these findings with the dual models constructed in the previous section. We find that   the metric \eqn{Kbeta} is equal to ${\mathcal{H}_{\tau,\alpha}}^{-1}$
%by dual transformation on the Hamiltonian of the $SU(2)$ PCM, 
in Eq. \eqn{Htaual-1}, for the following values of the parameters
\be\label{param}
 \beta= \pm  \frac{i}{2},\;\;\; \bar\tau=\pm i,   \,\,\, \alpha=\pm i.
\ee
%\textcolor{blue}{Penso che si possa ottenere il risultato $beta = 1/tau= 1\sqrt{2}$}.
In terms of the new metric one thus obtains:
\be
\tilde H_{ \beta}=\frac{1}{2}\int_{\mathbb{R}} \mathrm{d}\sigma \tI_I ({{\tilde{\mathcal{K}}}_{\beta}}^{-1})^{IJ} \tI_J   \label{sbdefoham}
\ee
which, for the choice of the parameters \eqn{param},  reproduces the Hamiltonian \eqn{hamww} that we have obtained by duality from the PCM $SU(2)$ model. 
\\
Notice however that, while the Poisson algebra of    the dual models constructed  in section \ref{PLD} is the full affine algebra of $\mathfrak{sl}(2,\C)$, here we only have a contraction of such an algebra, or in general a different algebra,  after rotating the fields with the B-transformation \eqn{oneparaB}.   
%we should not transform the Poisson algebra, as in section \ref{Born}, namely, we are not allowed to rotate the fields accordingly. In other words, the duality relation found  above between the $SU(2)$ PCM and the $SB(2,\C)$ PCM, cannot  be regarded as a $O(3,3)$ transformation of phase space coordinates. 
\\
Summarizing our findings, the natural $SB(2,\C)$ PCM model constructed in the Lagrangian approach,  has an Hamiltonian formulation given by the Hamiltonian \eqn{sbham} and the Poisson algebra \eqn{poidu1}-\eqn{poidu3}. 
%Alternatively, it can be described according to footnote [$\ref{note}$], where the Hamiltonian is the same as in Eq. \eqn{sbdefoham} and the Poisson algebra is represented by Eqs. \eqn{defo1}-\eqn{defo3}. 
On the other hand, the models which we have obtained in Section \ref{PLD} by performing a T-duality transformation of target space,  namely an $O(3,3)$ rotation, are described by the Hamiltonians \eqn{hamww} and Poisson algebra \eqn{tKtKcur}-\eqn{tItKcur}. 
The relation between the two, if any, is still unclear to us.  

\section{Double principal chiral model}\label{DPCM}

In the previous section we have succeeded in describing the Principal Chiral Modelof $SU(2)$ in terms of currents whose Poisson brackets furnish a realization of the affine algebra of the group $SL(2,\C)$, hence exhibiting a larger symmetry than the original Lagrangian approach.  Moreover, we have defined a natural model on the dual group of $SU(2)$ and we have exhibited a transformation which relates the Riemannian metrics of the two models.   It is therefore legitimate to look for a Lagrangian and an action with a manifest  $SL(2,\C)$ symmetry. 

\subsection{The Lagrangian formalism}
This is achieved by extending  the $SL(2,\C)$ action for the Isotropic Rigid Rotor reviewed in Section \ref{genacsec} to field theory. Hence, let us consider the group valued field:
$$ \Phi: \mathbb{R}^{1,1} \rightarrow \gamma  \in  SL(2,\mathbb{C})$$ 
and let us introduce the left-invariant  Maurer-Cartan one-form $\gamma^{-1}\mathrm{d}\gamma$
whose pull-back to to $\mathbb{R}^{1,1} $ reads as:
\be\label{mauca}
\Phi^*( \gamma^{-1}\mathrm{d}\gamma)=\gamma^{-1}\partial_t \gamma \mathrm{d}t+\gamma^{-1}\partial_{\sigma}\gamma \mathrm{d}\sigma
\ee
 which takes values in the Lie algebra $\mathfrak{sl}(2,\mathbb{C})$.  As previously, we shall not specify the pull-back from now on, unless necessary. Hence, upon using the Lie algebra basis $e_I=(e_i, \te^i)$ as in section \ref{genacsec}, one has:
\begin{align}
 \gamma^{-1}\partial_t \gamma= & \dot \bQ^I e_I, \\
 \gamma^{-1}\partial_{\sigma}\gamma= & {\bQ'^I}e_I.
\end{align}
with $\dot \bQ^I, \bQ'^I$, left generalized coordinates, respectively given by:
\be
 \dot \bQ^I=\Tr\left( \gamma^{-1}\partial_t \gamma e_I\right),\;\;\;    {\bQ'^I}=\Tr\left( \gamma^{-1}\partial_\sigma \gamma e_I\right)
 \ee
 with $\Tr$ the Cartan-Killing metric of $\mathfrak{sl}(2,\C)$. Moreover, as already done in Eqs. \eqn{AiBi}, we can use the product \eqn{prod1} to project the fiber coordinates along the bialgebra summands $\mathfrak{su}(2)$ and $\mathfrak{sb}(2,\C)$, according to 
 \beqa
 && \dot{Q}^i(\sigma,t) = 2{\rm Im} \Tr ( \gamma^{-1}\del_t\gamma \tilde e^i); \;\;\; \dot{\tilde{Q}}_i(\sigma,t)= 2{\rm Im} \Tr (\gamma^{-1}\del_t \gamma  e_i)\\
 &&Q'^i(\sigma,t) = 2{\rm Im} \Tr ( \gamma^{-1}\del_\sigma \gamma \tilde e^i); \;\;\; \tilde{Q}'_i(\sigma,t)= 2{\rm Im} \Tr (\gamma^{-1}\del_\sigma \gamma  e_i).
\eeqa
The Hodge operator applied to the Maurer-Cartan one-form  \eqn{mauca} exchanges the currents  $\dot \bQ^I$ and ${\bQ'^I}$ so to give:
\be
*_H \Phi^*[ \gamma^{-1}\mathrm{d}\gamma]=\gamma^{-1}\partial_t \gamma \mathrm{d}\sigma+\gamma^{-1}\partial_{\sigma}\gamma \mathrm{d}t  \,\,\, .
\ee
We therefore postulate the following action functional:
\be\label{PCMaction}
\mathcal{S}= \int_{\mathbb{R}^2} k_1 \braket{\Phi^*[\gamma^{-1} \dd \gamma] \wedge \h \Phi^*[\gamma^{-1} \dd \gamma]} +k_2 ((\Phi^*[\gamma^{-1} \dd \gamma]\wedge \h \Phi^*[\gamma^{-1} \dd \gamma]))
\ee
 which is the natural extension to field theory of the action introduced for the rigid rotor in  Eq.\eqn{genacIRR}.
 Upon introducing $k=k_1/k_2,$ the Lagrangian  is rewritten in terms of the left generalized coordinates 
%${\dot  Q}^I$
\be \label{lagrdoubl}
{\mathbf{L}}=   \frac{1}{2} \int_\R \dd\sigma\,  (k ~\eta+ \mathcal{H} )_{IJ} \left(\dot {\bQ}^I  \dot { \bQ}^J   -  {\bQ'^I } { \bQ'^J} \right)
 \ee    
 with 
\be
(k \ \eta+ \mathcal{H})_{IJ}= 
\begin{pmatrix}
 \delta_{ij} & k\delta_i^j+  \epsilon_i^{\,j3} \\
k \delta^i_j-\epsilon^i_{\,j3} & ( \delta^{ij}+ \epsilon^i_{k3} \epsilon^j_{l3}\delta^{kl})
\end{pmatrix}
.
\ee 
Recall that $\eta$ (Lorentzian) and $\mathcal{H}$ (Riemannian) are the left-invariant metrics on $SL(2, \mathbb{C})$ induced, respectively, by the pairings $2\mathrm{Im} \mathrm{Tr}()$ and $2\mathrm{Re}  \mathrm{Tr}()$ on $\mathfrak{sl}(2, \mathbb{C})$. They are two of the structures defining a Born geometry on $SL(2, \mathbb{C}).$ 
%\textcolor{red}{(Aggiungi commento su analogia Palatini-Holst e sottolinea differenza con articolo Demulder, Hassler, Thompson.)}
%Now, we have all the objects needed to state our action:
%\begin{equation}
%S=\frac{1}{2}\int_{\mathbb{R}^2} \bigl(\alpha \Braket{\gamma^{-1}\mathrm{d}\gamma,\gamma^{-1}\mathrm{d}\gamma}+ (\gamma^{-1}\mathrm{d}\gamma,\gamma^{-1}\mathrm{d}\gamma)_{\mathcal{R}}\bigr),
%\label{PCMaction}
%\end{equation}
%where the scalar products are given by \eqref{prod1} and \eqref{prod2}, i.e. $$\Braket{\gamma^{-1}\mathrm{d}\gamma ,\gamma^{-1}\mathrm{d}\gamma}=\bigl(I^I I^J\Braket{e_I,e_J}-J^IJ^J \Braket{e_I,e_J}\bigr)\mathrm{d}t \mathrm{d}\sigma$$ and $$(\gamma^{-1}\mathrm{d}\gamma,\gamma^{-1}\mathrm{d}\gamma)_{\mathcal{R}}=\bigl(I^I I^J(e_I,e_J)-J^IJ^J (e_I,e_J)\bigr)\mathrm{d}t \mathrm{d}\sigma. $$
%\subsection{Lagrangian and Hamiltonian dynamics of the Doubled model}

The Euler-Lagrange equation for the Lagrangian density \eqref{lagrdoubl} are:
\begin{equation}
\partial_t \frac{\partial \mathbf{L}}{\partial \dot{\bQ}^J}+ \partial_\sigma \frac{\partial \mathbf{L}}{\partial \bQ'^J }=(k\ \eta+\mathcal{H})_{IJ}\bigl(\partial_t\dot{\bQ}^J-\partial_{\sigma}\bQ'^J\bigr)= {\sf L}_{\bX_J}  \mathbf{L}
\end{equation}
with $\bX_J$ the left-invariant vector fields on the group manifold of $SL(2,\C)$. 
%Since the matrix $\alpha L_{IJ}-\mathcal{R}_{IJ}$ is non-vanishing, the Euler-Lagrange equation can be written as
%\begin{equation}
%\partial_t(\gamma^{-1}\partial_t\gamma)^I=\partial_{\sigma}(\gamma^{-1}\partial_{\sigma}\gamma)^I,
%\label{eullagr}
%\end{equation}
%thus, expliciting the components of the currents in the splitting of the Drinfel'd double, 
%\begin{equation}
%\partial_t I^i=\partial_{\sigma}J^i, \ \quad \ \partial_t\tilde{I}_i=\partial_{\sigma}\tilde{J}_i,
% \end{equation}
%we have the equations of motions of the two dual models. \\
Before passing to the Hamiltonian description, let us stress that the generalized action describes a kind of non-linear sigma model with target space $SL(2,\C)$, hence with doubled dimension with respect to the previous models. Because the model only contains the currents $\dot \bQ^J, \bQ'^J$, as previously we can read the latter as the tangent space coordinates of $TSL(2,\C)(\R)$.

\subsection{The Hamiltonian formalism} \label{hamifor}
According to the remark made at the end of the previous subsection, the Hamiltonian model will be interpreted as a model over the cotangent space $T^*SL(2,\C)(\R)$.  In order to obtain the Hamiltonian of the system,  the canonical momentum is computed:
\begin{equation} \label{mompi}
\bI_I=(I_i, \tilde{I}^i)=\frac{\delta\mathbf{L}}{\delta\dot{\bQ}^I}=(k\ \eta+\mathcal{H})_{IJ}\dot{\bQ}^J.  
\end{equation}
Let us recall that the matrix $(k \ \eta+ \mathcal{H})_{IJ}$ is invertible for $k^2 \neq 1$ and its inverse is
\be
[( k\, \eta + \mathcal{H})^{-1}]^{IJ}= \frac{1}{2} (1-k^2)^{-1} 
\begin{pmatrix}
\delta^{ij}+ \epsilon^i_{l3} \epsilon^j_{k3}\delta^{lk}& -{\epsilon^i}_{j3}-k \delta^i_j
\\
{\epsilon_i}^{j3}-k \delta_i^j& \delta_{ij} \,\,\, 
\end{pmatrix}  \,\,.    \nonumber
\ee
Therefore, the Legendre transform of \eqref{lagrdoubl}, obtained by inverting \eqref{mompi}, gives:
\begin{equation}
{\bf H}=\frac{1}{2}\int_{\mathbb{R}} \mathrm{d}\sigma \ \bigl([(k\, \eta+ \mathcal{H})^{-1}]^{IJ}\bI_I\bI_J+(k\, \eta+ \mathcal{H})_{IJ}{\bJ^I} {\bJ^J}\bigr).
\end{equation}
whereas we have for the Poisson brackets (see appendix \ref{appA})
\beqa 
\{\bI_I(\sigma'), \bI_J(\sigma'')\}&=& {C_{IJ}}^K \bI_K \delta(\sigma'-\sigma'') \label{poidoub1}\\
\{\bI_I(\sigma'), \bJ^J(\sigma'')\}&=& {C_{KI}}^J \bJ^K \delta(\sigma'-\sigma'')-\delta_I^J \delta'(\sigma'-\sigma'') \label{poidoub2}\\
\{\bJ^I(\sigma'), \bJ^J(\sigma'')\}&=& 0 \label{poidoub3}
\eeqa
and we have renamed $\bQ'^I\rightarrow \bJ^I$. The equations of motion read then as:
\beqa
\dot\bI_J&=&  \left\{\bI_M [ k\, \eta+ \mathcal{H})^{-1}]^{LK}\  \bI_K -  \bJ^L[ (k\, \eta+ \mathcal{H})^{-1}]_{LK}\  \bJ^K\right\} {C_{JL} }^M\nn\\
&+& \del_\sigma \bJ^L [(k\, \eta+ \mathcal{H})^{-1}]_{LJ} 
\eeqa

\subsection{Recovering the Chiral Model on $TSU(2)$}
Let us prove that we can recover the action on of the Principal Chiral Model on $TSU(2)$ with an appropriate gauging of the global symmetries of the generalized action.

Let us recall that 
\be
\Phi^*(\gamma \dd \gamma)= (\gamma^{-1} \del_t \gamma)^I e_I \dd t +  (\gamma^{-1} \del_\sigma \gamma)^I e_I \dd \sigma
\ee
can be projected along the two Lie algebras  according to  
\beqa
 \gamma^{-1} \del_t \gamma &=& A^i e_i + \tilde A_i \tilde e^i \\
  \gamma^{-1} \del_\sigma \gamma &=& B^i e_i + \tilde B_i \tilde e^i 
  \eeqa
  with 
  \beqa
  A^i&=& 2{\rm Im}\Tr  \gamma^{-1} \del_t \gamma \te^i\;\;\;\;  \tilde A_i= 2{\rm Im}\Tr  \gamma^{-1} \del_t \gamma te_i  \nn\\
  B^i&=& 2{\rm Im}\Tr  \gamma^{-1} \del_\sigma \gamma \te^i\;\;\;\;  \tilde B_i= 2{\rm Im}\Tr  \gamma^{-1} \del_\sigma \gamma te_i 
  \eeqa
We notice that, fixing the decomposition $\gamma=\tilde{g}g$, with $\tilde{g}\in SB(2,\mathbb{C})$ and $g \in SU(2)$, for any element $\gamma \in SL(2,\mathbb{C})$, the action \eqref{PCMaction} has manifest global symmetry under left action of $SB(2,\mathbb{C})$, called $SB(2,\mathbb{C})_L$, and $SU(2)_R$, the right action of $SU(2)$. 
We let the  $SB(2,\mathbb{C})_L$ symmetry become local, so we can introduce the connection one-form $C= C_i \te^i$ on the principal bundle $\R^{1,1} \times SB(2,\C) \rightarrow \R^{1,1}$ so that its pull-back (along any section) to $\R^{1,1}$ reads $C_i^t\te^i\mathrm{d}t+C_i^{\sigma}\te^i\mathrm{d}\sigma$, which takes values in the Lie algebra $\mathfrak{sb}(2,\mathbb{C})$. 
Hence we  modify the left-invariant one-form with the covariant derivative $\mathrm{D}=\mathrm{d}+C$:
\be
\Phi^*(\gamma^{-1} \mathrm{D} \gamma)= \Phi^*(\gamma^{-1} \dd \gamma) + \Phi^*(\gamma^{-1} C\gamma)=(\gamma^{-1} \del_t \gamma + \gamma^{-1} C^t\gamma) \dd t + (\gamma^{-1}  \del_\sigma \gamma + \gamma^{-1} C^\sigma \gamma) \dd t
\ee
and  define
\be
\gamma^{-1} \del_t \gamma + \gamma^{-1} C^t\gamma = U_i \te^i + W^i e_i
\ee
where 
\beqa
U_i&=&  A_i + C_j^t \,2{\rm Im}\Tr (\gamma^{-1} \te^j \gamma e_i) \\
W^i&=&  A^i + C_j^t \,2{\rm Im}\Tr (\gamma^{-1} \te^j \gamma \te^i)
\eeqa
and similarly 
\be
\gamma^{-1} \del_\sigma \gamma + \gamma^{-1} C^\sigma\gamma = V^i e_i + Z_i \te^i
\ee
with 
\beqa
V^i&=&  B^i + C_j^\sigma \,2{\rm Im}\Tr (\gamma^{-1} \te^j \gamma \te^i) \\
Z_i&=&  B_i + C_j^\sigma \,2{\rm Im}\Tr (\gamma^{-1} \te^j \gamma e_i).
\eeqa
In terms of the new degrees of freedom the generalized action \eqn{PCMaction}, with the gauge connection added, reads as:
\be
S_C=\frac{1}{2}\int_{\R^2}  \left[\delta_{ij} W^i W^j + 2(k \delta_i^j + \epsilon_i^{j3})W^i U_j + h^{ij} U_i U_j \right.\nn\\
-\left. \delta_{ij} V^i V^j + 2(k \delta_i^j + \epsilon_i^{j3})V^i Z_j + h^{ij} Z_i Z_j \right]
\ee
On  performing the following transformations
\beqa
\hat W^i &=& W^i + (k\delta^{ij} -\epsilon^{is3} )U_s \label{what}\\
\hat V^i &=& V^i + (k\delta^{ij} -\epsilon^{is3} )Z_s \label{vhat}
\eeqa
while $U_i, Z_i$ remaining unchanged, one gets for $S_C$ the following expression: 
\be
S_C=\frac{1}{2}\int_{\R^2}  \left[\delta_{ij} (\hat W^i \hat W^j  -\hat V^i \hat V^j) + (1-k^2) \delta^{ij} (U_i U_j- Z_i Z_j) \right]\dd\sigma \dd t %
\ee
The Wick-rotated generating functional of the gauged theory reads then as:
\begin{equation}
Z_C=\int \mathcal{D}g\mathcal{D}\tilde{g}\mathcal{D}C^t\mathcal{D}C^{\sigma} e^{-{S^{E}}_C},
\end{equation}
with $S^E_C$ the Euclidean gauge action, 
and we can trade the integration over $C^t, C^\sigma$ by an integration over the fields $U_i, Z_i$
\beqa\label{Zgen}
 Z_C&=&\int \mathcal{D}g\mathcal{D}\tilde{g} \ \det\biggl(\frac{\delta C^t_i}{\delta U_j}\biggr)\det\biggl(\frac{\delta C^{\sigma}_i}{\delta Z_j}\biggr)e^{-\frac{1}{2}\int_{\mathbb{R}^2}\mathrm{d}t\mathrm{d}\sigma \delta_{ij} (\hat W^i \hat W^j  +\hat V^i \hat V^j)}\nonumber \\ 
&\times& \int\mathcal{D}U_i \ e^{-\frac{1}{2}\int_{\mathbb{R}^2}\mathrm{d}t\mathrm{d}\sigma (1-k^2) \delta^{ij} U_i U_j} \int\mathcal{D}Z_i \ e^{-\frac{1}{2}\int_{\mathbb{R}^2}\mathrm{d}t\mathrm{d}\sigma (1-k^2) \delta^{ij} Z_i Z_j}.
\eeqa
For $-1\le k\le1$ the last two functional integrals can be performed yielding:
\be
\int\mathcal{D}U_i \ e^{-\frac{1}{2}\int_{\mathbb{R}^2}\mathrm{d}t\mathrm{d}\sigma (1-k^2) \delta^{ij} U_i U_j} = 
\int\mathcal{D}Z_i \ e^{-\frac{1}{2}\int_{\mathbb{R}^2}\mathrm{d}t\mathrm{d}\sigma (1-k^2) \delta^{ij} Z_i Z_j}=\left ( \frac{2\pi}{1-k^2}\right)^{\frac{3}{2}}
\ee
 Similarly, the Jacobian determinants  appearing in \eqn{Zgen} are constant, because the gauge transformation only involves constant matrices (see \cite{MPV18} for details). Therefore, up to a regularization factor which has to be introduced to take care  of the volume integration over the group $SB(2,\C)$,  we are left with 
\begin{equation} \label{partfin}
Z=\int \mathcal{D}g  \ e^{-\frac{1}{2}\int_{\mathbb{R}^2}\mathrm{d}t\mathrm{d}\sigma \delta_{ij} (\hat W^i \hat W^j  +\hat V^i \hat V^j)}.
\end{equation}
Upon observing that the transformations \eqn{what}, \eqn{vhat} give a redefinition of the fields $W^i,$ $V^i$ as $\hat W^i, \hat V^i$ still $\mathfrak{su}(2)$-valued, the partition function \eqref{partfin} clearly involves the Action of the Principal Chiral Model on the group $SU(2)$ 
%with the parameter $\alpha \in (-1,1)$ to preserve the sign of the Lagrangian
. Indeed, we can write the exponent of the derived partition function as 
\begin{equation}
S=\frac{1}{2}\int_{\mathbb{R}^2}\ Tr(g'^{-1}\mathrm{d}g'\wedge *g'^{-1}\mathrm{d}g'),
\end{equation}
with $g' \in SU(2),$ so we explicitly have the derivation of the model on $SU(2)$. Gauging the other symmetry, we obtain the model on $SB(2,\mathbb{C})$ as we discussed for the Isotropic Rigid Rotor, see \cite{MPV18} for details.\\

\section{Conclusions and Outlook}\label{conclus}
%Starting from an existing description of the dynamics of the Rigid Rotor on Heisenberg doubles \cite{MSS92}, we have introduced a new formulation, proposing a dual description of this model. To this, we have used the notion of Drinfel'dSummarizing our findings, the natural $SB(2,\C)$ PCM model  has an Hamiltonian formulation given by the Hamiltonian \eqn{sbham} and the Poisson algebra \eqn{poidu1}-\eqn{poidu3}. Alternatively, it can be described according to footnote [$\ref{note}$], where the Hamiltonian is the same as in Eq. \eqn{sbdefoham} and the Poisson algebra is represented by Eqs. \eqn{defo1}-\eqn{defo3}. 

%Moreover, we have shown that, from the "doubled"  Rotor, the usual description can be recovered gauging one of its symmetries. \\
%The discussion of the Rotor has been extended in order to describe Principal Chiral Models. To this, we have considered Double Geometry on Lie groups, using the notion of double group deriving from the Drinfel'd double, and Generalized Geometry, since these two formalisms coincide when built on Lie groups.\\

An alternative parametrization of the $SU(2)$ Principal Chiral Model  found  in ref. \cite{R89}, shows that the PCM, in its Hamiltonian formulation, can be given an equivalent description in terms of currents which span  a target phase space isomorphic to the group manifold of $SL(2,\C)$. Their Poisson algebra can be given the structure of the centrally extended  affine algebra $\mathfrak{sl}(2,\C)(\R)$.  Following a previous paper of the authors, \cite{MPV18}, the model is here studied  as a higher dimensional generalization of the Isotropic Rigid Rotor dynamics  with the aim of further deepening its remarkable geometric structures. 
\\
The standard Hamiltonian formulation of the $SU(2)$ PCM model exploits the fact that the dynamics is fully described by fields, the currents, which  span $T^*SU(2)$ as target phase space and act as infinitesimal generators of an affine algebra which is the semi-direct sum  $\mathfrak{su}(2)(\R)\dot\oplus \mathfrak{a} (\R)$. We speculate on the fact that, as a Lie group, $T^*SU(2)$ is  the trivial Drinfel'd double   of the group $SU(2)$, which we have called  the classical double. The latter  gives rise to a fully nontrivial Drinfel'd double, the group $SL(2,\C)$, when the Abelian subalgebra of the semidirect sum is deformed to that of $SB(2,\C)$. By exploiting this property, we first review in detail the derivation of a whole family of  equivalent PCM models described in terms of  current algebra of the group $SL(2,\C)$, we thus   show that they can actually be interpreted in terms of Born geometries related by B-transformations. We then  perform $O(3,3)$ transformations of such  a family and find a parametric family of T-dual PCM models, with target configuration space the group $SB(2,\C)$,   the Poisson-Lie dual  of $SU(2)$ in the Iwasawa decomposition of the Drinfel'd double $SL(2, \mathbb{C} )$.  Poisson-Lie symmetries are discussed. 
%A relevant result obtained in this context is that the PCM, in the formulation given by one of the equivalent Hamiltonians, in particular the one in Eq. \eqn{modiha3}, together with the Poisson algebra \eqn{IIcur}-\eqn{IKcur}, is a Poisson-Lie sigma-model.
Then, a natural Lagrangian model has been constructed directly on the dual group $SB(2,\C)$.  Its relation to the dual models previously introduced is still unclear to us and needs further analysis.   
Finally we have introduced a double PCM with the group manifold of $SL(2,\C)$ as its target configuration space and $TSL(2,\C)$ as the target tangent  space. The degrees of freedom are thus doubled. We have shown, performing a gauging of its symmetries, that both the Lagrangian models, with  $SU(2)$ and $SB(2,\C)$ target configuration spaces, can be retrieved. 
 %The natural $SB(2,\C)$ PCM model  constructed has an Hamiltonian formulation given by the Hamiltonian \eqn{sbham} and the Poisson algebra \eqn{poidu1}-\eqn{poidu3}. Alternatively, it can be described %according to footnote [$\ref{note}$], 
 %by using the Hamiltonian Eq. \eqn{sbdefoham} and with the Poisson algebra represented by Eqs. \eqn{defo1}-\eqn{defo3}. 
%An interesting new kind of duality has emerged out in this context:  the models that have been obtained  by performing a T-duality transformation of target space,  namely an $O(3,3)$ rotation, are described by either of  those two particular Hamiltonian functions but with Poisson algebras exchanged. The meaning of this new duality is still under investigation.

 %Finally, in order to build a model where the symmetries exhibited by the dynamics are manifest, a parent action is constructed with target configuration space  the Drinfel'd double $SL(2,\C)$, hence doubling the degrees of freedom.   From it, either of the dual partner models can be recovered, by gauging one of its global symmetries. 

A further extension of this model can be given adding a Wess-Zumino term \cite{BPV}. This could provide a deeper insight, among other things,  on the geometric structures  of String Theory on $AdS_3$, the study of which is interesting from the point of view of the $AdS/CFT$ correspondence since it enables to study the correspondence beyond the gravity approximation \cite{MO, MO2,MO3}.

Last but not least, all what we have learnt from this model could be further extended to the world-sheet string action. In this case, a manifestly $O(d,d)$-invariant action may be written, considering that the configuration space is no longer a Lie group, but a differentiable manifold.  It would be interesting to follow this way, in which $O(d,d)$-invariance is implemented writing a doubled string action, as discussed for Principal Chiral Models, and then performing the low energy limit. This limit  result should reproduce all the results so far obtained in Double Field Theory.

\noindent{\bf Acknowledgements} 
 P.V.  acknowledges  support by COST (European Cooperation in Science  and  Technology)  in  the  framework  of  COST  Action  MP1405  QSPACE. V.E.M. thanks Richard Szabo for helpful discussions. The work of V.E.M. was funded by the Doctoral Training Grant ST/R504774/1 from the UK Science and Technology Facilities Council (STFC). F.P. thanks the Simon Center for Geometry and Physics for their hospitality and support during the Simons Summer Workshop 2018. 
 
 \appendix
 \section{Appendix: Poisson brackets}\label{appA}
In this appendix we derive  the  current algebras \eqn{IIbr}-\eqn{JJbr},  \eqn{poidu1}-\eqn{poidu3}, \eqn{poidoub1}-\eqn{poidoub3}  from the canonical one-form obtained by the relevant action functional.
 
 Let us start with the standard formulation of the principal $SU(2)$ chiral model, whose action is given by \eqn{act1}. 
 As for the rigid rotor, we choose the parametrization 
 \be
 \phi: \sigma\in\R\rightarrow g(\sigma)= 2(y^0(\sigma) e_0+ i y^i(\sigma) e_i)
 \ee
 with $\sum_\mu  y^\mu y^\mu= 1$
 Upon defining $I= -\frac{i}{2} I_i {e^i}^*$ with ${e^i}^*(e_j)= \delta^i_j$, and recalling that $g^{-1} \dd g= 2i \alpha^k e_k$, we have for the canonical one-form 
 \be
 \Theta= \int_{\R} <I | g^{-1} \dd g>= \int_{\R}  I_i (\sigma) \alpha^i(\sigma)
 \ee
 so that
 \be
 \Omega = \int_{\R}   \dd I_i (\sigma)\wedge  \alpha^i (\sigma) + I_i(\sigma){\epsilon_{jk}}^i \alpha^j(\sigma)\wedge  \alpha^k(\sigma)
% &=&  \int_{\R} d A^i \wedge\delta_{ij} (d A^j  + d J^j )+ \epsilon_{ikl}A^i   (d A^k + dJ^k) \wedge 
% (d A^l + dJ^l )
\ee
%Hamiltonian vector fields:
%\be
%Y_{I_j}: i_{Y_{I_j}}\omega= d I_j \Rightarrow \;\; Y_{I_j}= \frac{\delta}{\delta J^j(\sigma)}
%\ee
with $\alpha^i(\sigma) = [y^0 \dd y^i -y^i \dd y^0 + {\epsilon_{jk}}^i y^j \dd y^k ](\sigma) $ the left-invariant one-forms on the group manifold, in the chosen parametrization.
The Poisson structure is thus
\be
\Lambda=  \int_{\R} \dd\sigma\, \left(X_i(\sigma) \wedge \frac{\delta}{\delta I_i (\sigma)} + {\epsilon_{jk}}^i I_i  \frac{\delta}{\delta I_j (\sigma)} \wedge \frac{\delta}{\delta I_k (\sigma)}\right)
 \ee
 with $X_i(\sigma) $ the left-invariant vector fields which are dual to the one-forms $\alpha^i(\sigma)$, that is, in the chosen parametrization
 \be
 X_i(\sigma) = y^0 \frac{\delta}{\delta y^i(\sigma)}  -y^i \frac{\delta}{\delta y^0(\sigma)} + {\epsilon_{ij}}^k y^j(\sigma) \frac{\delta}{\delta y^k(\sigma)} 
 \ee
 We thus obtain
 \beqa\label{fundpb}
 \{I_i(\sigma'), I_j{\sigma''}\}&=& {\epsilon_{ij}}^k I_k(\sigma') \delta(\sigma'-\sigma'')\nonumber\\
 \{y^i(\sigma' ),I_j(\sigma'')\}&=&[\delta^{i}_j y^0(\sigma') + {\epsilon_{jk}}^i\,y^k(\sigma')] \delta (\sigma'-\sigma'')\;\;\;\;\; {\rm or  }\;\; \{g(\sigma'), I_j(\sigma'')\}= 2  i  g(\sigma') e_j \delta(\sigma'-\sigma'') \nonumber\\
 \{y^0(\sigma' ),I_j(\sigma'')\}&=&-y^j(\sigma') \delta (\sigma'-\sigma'')\nonumber\\
  \{y^\mu(\sigma' ),y^\nu(\sigma'')\}&=& 0 \;\;\;\;\;{\rm or ~ } \;\; \{g(\sigma'), g(\sigma'')\}=0 
 \eeqa
On using $J^i(\sigma)=-i \Tr (g^{-1} \del_\sigma g) e_i = y^0 \del_\sigma y^i -y^i \del_\sigma y^0 + {\epsilon_{jk}}^i y^j \del_\sigma y^k  $  we compute 
\beqa
 \{J^i(\sigma'), I_j({\sigma''})\}&=& \Tr e_i \{g^{-1} \del_{\sigma' } g, I_j(\sigma'')\} \nonumber\\
 &= &-i \Tr e_i [-g^{-1}\{g(\sigma'), I_j(\sigma'')\} g^{-1}\del_{\sigma'} g + g^{-1}\{\del_{\sigma'} g, I_j(\sigma'')\}]
 \eeqa
 which can be seen to give \eqn{mixbrac} because of the second of the brackets \eqn{fundpb}. Analogously we can compute 
 \be
 \{J^i(\sigma'),J^j{(\sigma'')}\}= \{\Tr e_i g^{-1} \del_{\sigma' } g,\Tr e_j g^{-1} \del_{\sigma'' } g\}=0
 % \nonumber\\&= &\Tr e_i [-g^{-1}(\sigma')\{g(\sigma'), I_j(\sigma''\} g^{-1}(\sigma')\del_{\sigma'} g + g^{-1}\{\del_{\sigma'} g, I_j\}]
 \ee
 because group variables have zero Poisson brackets according to  the last of Eqs. \eqn{fundpb}.
 
 An analogous computation can be performed for the Poisson brackets of the chiral model on the Poisson-Lie dual group $SB(2,\C)$.
The action functional for the model is represented by  \eqn{dualchiact}. 
 As in section \ref{DIRR}  we choose the parametrization 
 \be
 \tilde{\phi}: \sigma\in\R\rightarrow \tg(\sigma)= 2(u^0(\sigma) \te^0+ i u^i(\sigma) \te^i)
 \ee
 with $(u^0)^2-(u^3)^2= 1$. On introducing $\tI= -i \tI^i \te_i^*$, with $\te_i^*(\te^j)=\delta_i^j$ and recalling that $\tg^{-1} \dd\tg= i \talpha_j\te^j$
 We have for the canonical one-form 
 \be
 \tilde{\Theta}= \int_{\R} <\tI | \tg^{-1} \dd \tg>= \int_{\R}  \tI^i (\sigma) \talpha_i
 \ee
 so that
 \be
\tilde \Omega = \int_{\R}   \dd\tI^i (\sigma)\wedge  \talpha_i (\sigma) + \tI^i  (\sigma) {f^{jk}}_i \talpha_j (\sigma) \wedge  \talpha_k (\sigma) 
% &=&  \int_{\R} d A^i \wedge\delta_{ij} (d A^j  + d J^j )+ \epsilon_{ikl}A^i   (d A^k + dJ^k) \wedge 
% (d A^l + dJ^l )
\ee
%Hamiltonian vector fields:
%\be
%Y_{I_j}: i_{Y_{I_j}}\omega= d I_j \Rightarrow \;\; Y_{I_j}= \frac{\delta}{\delta J^j(\sigma)}
%\ee
with $\talpha_i(\sigma) = 2 [u^0 \dd u^i -u^i \dd u^0 + {f^{jk}}_i u_j \dd u_k ](\sigma) $ the left-invariant one-forms on the group manifold, in the chosen parametrization.
The Poisson structure is thus
\be
\tilde{\Lambda}=  \int_{\R} \tX^i(\sigma) \wedge \frac{\delta}{\delta \tI^i (\sigma)} + {f^{jk}}_i \tI^i
 \frac{\delta}{\delta \tI^j (\sigma)} \wedge \frac{\delta}{\delta \tI^k (\sigma)}
 \ee
 with $\tX^i(\sigma) $ the left-invariant vector fields which are dual to the one-forms, that is, in the chosen parametrization
 \be
 \tX^i(\sigma) = \frac{1}{2} \left(u^0 \frac{\delta}{\delta u^i(\sigma)}  -u^i \frac{\delta}{\delta u^0(\sigma)} - {f^{ik}}_j u^j(\sigma) \frac{\delta}{\delta u^k(\sigma)} \right)
 \ee
 We thus obtain
 \beqa\label{funduapb}
 \{\tI^i(\sigma'), \tI^j{\sigma''}\}&=& {f^{ij}}_k\tI^k(\sigma') \delta(\sigma'-\sigma'')\nonumber\\
 \{u^i(\sigma' ),\tI^j(\sigma'')\}&=&\frac{1}{2}[\delta^{ij} u^0(\sigma') + {f^{ij}}_k u^k(\sigma')] \delta (\sigma'-\sigma'')\;\;\;\;\; {\rm or  }\;\; \{\tg(\sigma'), \tI^j(\sigma'')\}=  2  \tg(\sigma') \te^j \delta(\sigma'-\sigma'') \nonumber\\
 \{u^0(\sigma' ),\tI^j(\sigma'')\}&=&-\frac{1}{2} u^j(\sigma') \delta (\sigma'-\sigma'')\nonumber\\
  \{u^\mu(\sigma' ),u^\nu(\sigma'')\}&=& 0 \;\;\;\;\;{\rm or ~ } \;\; \{\tg(\sigma'), \tg(\sigma'')\}=0 
 \eeqa
% \textcolor{red}{rivisto FINO A QUA}
 
On using $\tJ_i(\sigma)= \langle \tg^{-1} \del_\sigma \tg,  e_i\rangle = u^0 \del_\sigma u^i -u^i \del_\sigma u^0 +  {f^{ik}}_{j}u^j \del_\sigma u^k  $  we compute 
\beqa
 \{\tJ_i(\sigma'), \tI^j{\sigma''}\}&=& 2Im \Tr e_i \{\tg^{-1} \del_{\sigma' } \tg, \tI^j(\sigma'')\} \nonumber\\
 &= &2 Im \Tr e_i [-\tg^{-1}(\sigma')\{\tg(\sigma'), \tI^j(\sigma''\} \tg^{-1}(\sigma')\del_{\sigma'} \tg + \tg^{-1}\{\del_{\sigma'} \tg, \tI^j\}]
 \eeqa
 which can be seen to give \eqn{poidu2} because of the second of the brackets \eqn{funduapb}. Similarly we can compute 
 \be
 \{\tJ^i(\sigma'),\tJ^j{\sigma''}\}= \{2Im\Tr e_i \tg^{-1} \del_{\sigma' } \tg,2Im\Tr e_j \tg^{-1} \del_{\sigma'' } \tg\}
 % \nonumber\\&= &\Tr e_i [-g^{-1}(\sigma')\{g(\sigma'), I_j(\sigma''\} g^{-1}(\sigma')\del_{\sigma'} g + g^{-1}\{\del_{\sigma'} g, I_j\}]
 \ee
 where the latter is zero because group variables have zero Poisson brackets according to last of Eqs. \eqn{funduapb}.
 
 Finally, we derive the Poisson brackets \eqn{poidoub1}-\eqn{poidoub3} for the $\mathfrak{sl}(2,\C)(\R)$ current algebra. 
 Upon defining $\bI= -\frac{1}{2} \bI_I {e^I}^*$ with ${e^I}^*(e_I)= \delta^I_J$, and recalling that $\gamma^{-1} \dd \gamma= 2 \zeta^K e_K$, with $\zeta^K$ the $SL(2,\C)$ left-invariant one-forms, we have for the canonical one-form 
 \be
 \Theta_D= \int_{\R} <\bI | \gamma^{-1} \dd \gamma>= \int_{\R}  \bI_I (\sigma) \zeta^I(\sigma)
 \ee
 so that
 \be
 \Omega_D = \int_{\R}   \dd\bI_I (\sigma)\wedge  \zeta^I (\sigma) + \bI_I(\sigma){C_{JK}}^I \zeta^J(\sigma)\wedge  \zeta^K(\sigma)
 \ee
 The Poisson structure is thus
\be
\Lambda_D=  \int_{\R} \dd\sigma\, \left(\bX_I(\sigma) \wedge \frac{\delta}{\delta \bI_I (\sigma)} + {C_{JK}}^I \bI_I(\sigma)  \frac{\delta}{\delta \bI_J (\sigma)} \wedge \frac{\delta}{\delta \bI_K(\sigma)}\right)
 \ee
 with $\bX_I(\sigma) $ the left-invariant vector fields which are dual to the one-forms $\zeta^I(\sigma)$. 
 %, that is, in the chosen parametrization
 %\be
 %X_i(\sigma) = y^0 \frac{\delta}{\delta y^i(\sigma)}  -y^i \frac{\delta}{\delta y^0(\sigma)} + {\epsilon_i}_{jk}y^j(\sigma) \frac{\delta}{\delta y^k(\sigma)} 
 %\ee
 %We thus obtain
 
{We thus compute the Poisson brackets. For the sake of simplicity, we do not choose any parametrization for $SL(2, \mathbb{C}).$ A similar computation can be analogously carried on for the Poisson brackets on $SU(2)$ and $SB(2, \mathbb{C}),$ since we always deal with matrix Lie groups.   
 The first Poisson bracket is straightforward
 \be\label{ffundpb1}
 \{\bI_I(\sigma\rq{}), \bI_J(\sigma\rq{}\rq{})\}=\Lambda_D(\mathrm{d} \bI_I(\sigma\rq{}), \mathrm{d} \bI_J(\sigma\rq{}\rq{}) )= {C_{IJ}}^K \bI_K(\sigma\rq{}) \delta(\sigma\rq{}-\sigma\rq{}\rq{}) \, .
 \ee
In order to derive the remaining brackets, we  compute 
 \begin{align} 
  \{\gamma(\sigma\rq{}), \bI_J(\sigma\rq{}\rq{})\}= &\Lambda_D(\mathrm{d} \gamma(\sigma\rq{}),\mathrm{d} \bI_J(\sigma\rq{}\rq{})) \nonumber \\
  = & \bX_J(\sigma\rq{}\rq{}) (\gamma(\sigma\rq{}) \gamma^{-1}(\sigma\rq{}) \mathrm{d} \gamma(\sigma\rq{}))  \label{ffundpb2} \\
  = & 2 \gamma(\sigma') e_J \delta(\sigma\rq{}-\sigma\rq{}\rq{}) \nonumber .
  \end{align}
  Notice that we could have performed the same calculation for the groups  $SU(2)$ and $SB(2,\C)$ where the analogous result was instead obtained by choosing explicitly a parametrization.  The above calculation can be carried on for any matrix Lie group. 
Finally, 
  \be \label{ffundpb3}
 \{\gamma(\sigma'), \gamma(\sigma'')\}=0 \,
 \ee
  because there are no terms in $\Lambda_D$ involving the wedge product of two left-invariant vector fields. 
On using $\bJ^I(\sigma)=\Tr (\gamma^{-1} \del_\sigma \gamma) e_I ,$  we compute 
\beqa
 \{\bJ^I(\sigma'), \bI_J({\sigma''})\}&=& \Tr e_I \{\gamma^{-1} \del_{\sigma' } \gamma, \bI_J(\sigma'')\} \nonumber\\
 &= &\Tr e_I [-\gamma^{-1}\{\gamma(\sigma'), \bI_J(\sigma'')\} \gamma^{-1}\del_{\sigma'} \gamma + \gamma^{-1}\{\del_{\sigma'} \gamma, \bI_J(\sigma'')\}]
 \eeqa
 which can be seen to give \eqn{poidoub2} because of the Poisson brackets \eqn{ffundpb2}. Analogously we can compute 
 \be
 \{\bJ^I(\sigma'),\bJ^J{(\sigma'')}\}= \{\Tr e_I \gamma^{-1} \del_{\sigma' } \gamma,\Tr e_J \gamma^{-1} \del_{\sigma'' } \gamma\}=0 \, ,
 % \nonumber\\&= &\Tr e_i [-g^{-1}(\sigma')\{g(\sigma'), I_j(\sigma''\} g^{-1}(\sigma')\del_{\sigma'} g + g^{-1}\{\del_{\sigma'} g, I_j\}]
 \ee
which gives \eqn{poidoub3} because group variables have zero Poisson brackets according to Eq. \eqn{ffundpb3}.

\end{document}